\documentclass[%
 aip,
 cha,
 amsmath,amssymb,
 reprint,%
floatfix]{revtex4-1}

\usepackage[a4paper]{geometry}
\usepackage{natbib}
\usepackage[utf8]{inputenc}
\usepackage{float}
\usepackage{rotating}
\usepackage[labelfont=bf, justification=centering, labelsep=colon]{caption}
\usepackage[labelformat=parens, labelsep=space, justification=centering]{subcaption}
\renewcommand{\thefigure}{\arabic{figure}}

\let\rho\varrho

\usepackage{bbold}
\usepackage{nicefrac}

\expandafter\let\csname equation*\endcsname\relax
\expandafter\let\csname endequation*\endcsname\relax
\usepackage{amsmath}

\usepackage{graphicx}
\usepackage{wrapfig}
\usepackage[export]{adjustbox}

\usepackage{amsfonts}
\newcommand{\overbar}[1]{\mkern 1.5mu\overline{\mkern-1.5mu#1\mkern-0mu}\mkern 0mu}

\usepackage{xcolor}
\definecolor{darkblue}{rgb}{0.2, 0.2, 0.6}
\usepackage[colorlinks, linkcolor=black, citecolor=darkblue, filecolor=black, urlcolor=darkblue]{hyperref}

\makeatletter
\def\@email#1#2{%
 \endgroup
 \patchcmd{\titleblock@produce}
  {\frontmatter@RRAPformat}
  {\frontmatter@RRAPformat{\produce@RRAP{*#1\href{mailto:#2}{#2}}}\frontmatter@RRAPformat}
  {}{}
}%
\makeatother
\begin{document}








\title[Resilience basins of complex systems: Prosumer impacts on power grids]{Resilience basins of complex systems: an application to prosumer impacts on power grids}

\author{Samuel Bien}
\affiliation{Institute of Environmental Science and Geography, Potsdam University, 14469 Potsdam, Germany}
\affiliation{Institute of Physics, Potsdam University, 14469 Potsdam, Germany}
\affiliation{Complexity Science Department, Potsdam Institute for Climate Impact Research, 14473 Potsdam, Germany}
\affiliation{FutureLab for Earth Resilience in the Anthropocene, Earth System Analysis Department, Potsdam Institute for Climate Impact Research, 14473 Potsdam, Germany}

\author{Paul Schultz}
\affiliation{50Hertz Transmission GmbH, 10557 Berlin, Germany}

\author{Jobst Heitzig}
\affiliation{FutureLab for Game Theory and Networks of Interacting Agents, Potsdam Institute for Climate Impact Research, 14473 Potsdam, Germany}

\author{Jonathan F. Donges}
\affiliation{FutureLab for Earth Resilience in the Anthropocene, Earth System Analysis Department, Potsdam Institute for Climate Impact Research, 14473 Potsdam, Germany}
\affiliation{Stockholm Resilience Centre, Stockholm University, 106 91 Stockholm, Sweden}




\date{\today}

\begin{abstract}
Comparable to the traditional notion of stability in system dynamics, resilience is typically 
measured in a way that assesses the quality of a system's response, for example the speed of its recovery.
We present a broadly applicable complementary measurement framework that quantifies resilience similarly to basin stability by estimating a \emph{resilience basin} which reflects the extent of adverse influences that the system can recover from in a sufficient manner. In contrast to basin stability, the adverse influences considered here are not necessarily displacements in state space, but arbitrarily complex impacts to the system, quantified by adequate parameters.

As a proof of concept, we present two applications: (i) the well-studied single-node power system as an easy-to-follow example and (ii) a stochastic model of a low-voltage DC power grid undergoing an unregulated energy transition consisting in the random appearance of prosumers. These act as decentral suppliers of photovoltaic power and alter the flow patterns while the grid topology remains unchanged.
The resilience measurement framework is applied to evaluate the effect and efficiency of two response options:
(i) upgrading the capacity of existing power lines 
and (ii) installing batteries in the prosumer households.

The framework demonstrates that line upgrades can provide potentially unlimited resilience against energy decentralization, while household batteries are inherently limited (achieving $\leq70$\% of the resilience of line upgrades). 
Further, the framework aids in optimizing budget efficiency by pointing toward threshold budget values as well as budget-dependent ideal strategies for the allocation of line upgrades and for the battery charging algorithm.
\end{abstract}


\maketitle


\textbf{\emph{Resilience} is a term used across many different scientific disciplines to describe the property of a system to remain functional during (or after) external adverse influences (a.k.a. shocks, perturbations, disruptions, etc.). It stems from the Latin verb \emph{resilire} which literally means \emph{to jump back}. Despite its popularity, there is no universal definition of how to measure resilience. Rather, there are many different approaches depending on the discipline and the system.\\
A related term, \emph{stability}, is a more narrow and more well-defined concept and describes how a system's actual internal state (not its functionality) reacts to perturbations. In this paper, we use a specific stability measure called \emph{basin stability} as inspiration for a novel framework of how to quantify resilience. Our measure, which we call \emph{basin resilience}, is defined in a very general way and can be applied to essentially any context: Crucially, it is flexible regarding the modeler's concept of what the essence of resilient behavior is. Further, it can be applied not only to dynamical systems governed by differential equations, but also to e.g. time-discrete and stochastic systems.\\
After defining the mathematical framework of basin resilience, we demonstrate its relation to basin stability by applying it to a well-studied conceptual model of a single generator node connected to a power grid.\\
We then apply our framework to a simplistic power grid model. In the model, we assume DC power flows (like water being distributed through a pipe network) and we evaluate the impact of randomly appearing \emph{prosumers}: consumers who also produce a certain amount of power themselves, in our case by photovoltaics. These unforeseen decentral sources of power threaten to overload existing power lines, causing line shutdowns and potentially local blackouts. Using our framework, we evaluate two response options that the power grid operator could implement in reaction to these prosumers: (i) Increasing the flow capacities of certain vulnerable power lines and (ii) installing batteries in the prosumer households to act as buffers for dangerous fluctuations in the solar power supply.\\
Our simple model application does not produce novel insights about power grid dynamics, and it also does not allow for real-world implications of how to improve power grid resilience. But it demonstrates that our framework can uniquely quantify and therefore compare resilience based on arbitrarily complex adverse influences and response strategies. For example, battery installation achieves at most 70\,\% of the resilience that the best line upgrades can provide in our scenario. In addition, we evaluate the resilience depending on the response budget (e.g. the total capacity upgrades on all lines) and other parameters (e.g. the spatial allocation scheme of the upgrades). This demonstrates the potential to inform policy choices by determining the optimal benefit-to-cost ratio.}

\section{Introduction}

Due to the continuously increasing role of technologies in all aspects of human civilization, an important subject of research is the dynamics of complex social-technological systems \citep{technosphere}.
They are particularly relevant in the context of climate change, since its mitigation requires both technological innovation and societal transformation.

Meanwhile, the notion of resilience, originally introduced in the context of ecology by \citet{holling}, is becoming more and more popular throughout a wide range of disciplines. Examples include socio-ecology \citep{folke2010}, economics \citep{rose}, and engineering \citep{whitson}.
Across those disciplines, different definitions are used, and even within disciplines, there exists a multitude of approaches to measuring resilience. An overview of different possible resilience measures is given by \citet{ingrisch}.

This paper does not aim to unify these different measurement approaches, but instead presents a way to build on them and create a new perspective on measuring resilience: We argue that most systems are vulnerable not only to a single type of threat, but to a range of threats, which can vary in terms of both their probability of occurrence and their severity. Therefore, confident planning of resilient systems requires a rigorous, complete resilience analysis.

Our framework is not limited to social-technological systems but can, in principle, be applied to systems of any category. It is inspired by the concept of basin stability which was introduced by \citet{menck} to quantify the stability of attractors in multi-stable dynamical systems, e.g., power grids \citep{menck2014dead,mitra2017multiple}.
Previously, an attractor's stability was commonly measured locally, for example by evaluating the curvature of the potential landscape at an equilibrium point, indicating the magnitude of the restoring force.
In contrast, basin stability evaluates an attractor non-locally by measuring the phase space volume of its basin of attraction, which is the set of all states from which the system will evolve back to the attractor.

Similarly, existing resilience measures tend to focus on \emph{how well} the system can respond to a specific adverse influence, evaluating, for example, the recovery speed. Instead, our resilience measurement framework quantifies \emph{how often} or \emph{how likely} the system's response is \emph{good enough}, considering all possible influences.
We call the set of influences that the system can cope with its \emph{resilience basin}.\\

\noindent One relevant example of 
complex systems is networks, particularly infrastructure networks such as power grids. Several authors have studied the resilience of diverse types of networks, e.g., computer networks \citep{najjar1990network}, communication networks \citep{smith2011network}, or collaboration networks \citep{liu2011attack}, to mention just a few, and also somewhat more generally flow networks \citep{kaiser2021topological}. Also, the resilience of power grids has already being studied extensively, typically regarding extreme weather events or other external influences such as cyber-attacks \citep{bie, jufri, braun2020blackouts, schafer2018dynamically}. However, the concept of power grid resilience is also useful in the context of the necessary shift toward renewable energies, for example the German Energiewende. Generally speaking, the study of resilience does not have to (and we argue, it should not) be applied exclusively to uncontrollable, external threats to a system. Instead, it is also sensible for examining detrimental side effects of intentional processes.\\

\noindent This paper is structured as follows: In Sec. \ref{sec:definitions}, we clarify the terminology regarding resilience that we use throughout the paper, and we define our resilience measurement framework.
In Sec. \ref{sec:application_node}, we apply our framework to the well-known simple dynamical system of a single power grid node, demonstrating its relation to basin stability.
In Sec. \ref{sec:application_grid}, we then introduce a more complex, time-discrete probabilistic power grid model and explain how we apply our measurement framework to it.
In Secs. \ref{sec:results} and \ref{sec:discussion}, we analyze and discuss the results of the exemplary applications.
Finally, the paper ends with our conclusion in Sec. \ref{sec:conclusion}.

\section{Definitions}
\label{sec:definitions}

\subsection{Resilience Terminology}

Even though the notion of resilience has been established in the scientific literature for half a century, its exact definition is still being discussed. Especially the relation between resilience and stability is subject to an ongoing debate \citep{van2021unifying}.

In an attempt to standardize the terminology of resilience, \citet{tamberg} proposed a general systematic framework for describing and specifying resilience of 
complex systems. It consists of four parts:
\begin{enumerate}
    \item \emph{Resilience of what?} One has to specify which system---or family of systems---is the subject of analysis. It has to be clear which are its internal mechanisms, as well as interactions to external entities. Relevant aspects are also uncertainties and vulnerabilities of the system, if known.
    \item \emph{Resilience regarding what?} There needs to be a quantifiable measure of what the system essentially aims to achieve, as subjectively defined by the modeler. This quantity is typically called the system performance or system functionality. \citet{tamberg} call it the \emph{sustainant}: it is the quantity that has to be sustained. Further, it has to be specified which sustainant value (or range of values) is \emph{acceptable} for the system. It is important to note that the sustainant is not the same as the (micro-)state of the system. A system can have the same sustainant value while being in different states, as also pointed out by \citet{schoenmakers2021resilience}.
    \item \emph{Resilience against what?} Of all the possible events that can impact the sustainant, it has to be specified which ones are examined. These events may be called perturbations, disturbances or disruptions. \citet{tamberg} refer to them as the \emph{adverse influence} of the system, or \emph{influence} as shorthand.
    \item \emph{Resilience by what means?} It has to be clarified whether the system passively restores itself, or if it can actively oppose the adverse influence---and if so, how. \citet{tamberg} call the available means of the system its \emph{response options}.
\end{enumerate}

\noindent Resilient systems can be classified by their response options:
\citet{schoenmakers2021resilience} separate them into (i) tolerance and flexibility and (ii) adaptation and transformation.
Similarly, \citet{barfuss} differentiate between \emph{persistence} resilience, \emph{adaptation} resilience, and \emph{transformation} resilience. Persistence/tolerance/flexibility resilience is the most basic and widespread notion of resilience, reflecting the \emph{passive} ability of a system to manage an adverse influence without changing itself. We argue that this actually makes this type of resilience equivalent to stability or elasticity.

In contrast, the other types of resilience implicate some kind of change that the system undergoes to \emph{actively} recover from the adverse influence: Adaptation can be understood as \emph{quantitative} changes to existing parameters within the system, while transformation means change on a \emph{qualitative} level, for example the addition or removal of certain system parts.
Thus, we argue that control theory is actually a study of how to achieve adaptive resilience: A system is engineered to check itself for failures (sustainant deficits) and respond to them by the means of varying the system parameters within pre-existing bounds.

Within the many interpretations of resilience, there are generally two aspects of it which are recognized universally, as thoroughly discussed by \citet{hodgson2015you}: On the one hand, \emph{resistance} against the decline of the sustainant, and on the other hand,\emph{recovery} from low sustainant values back into the acceptable range. Some disciplines have historically focused their notion of resilience on only one of these aspects, which is why \citet{ingrisch} refer to resistance-centered concepts as \emph{ecological resilience} and to those focused on recovery as \emph{engineering resilience}.

\begin{figure*}
    \centering
    \includegraphics[width=\linewidth]{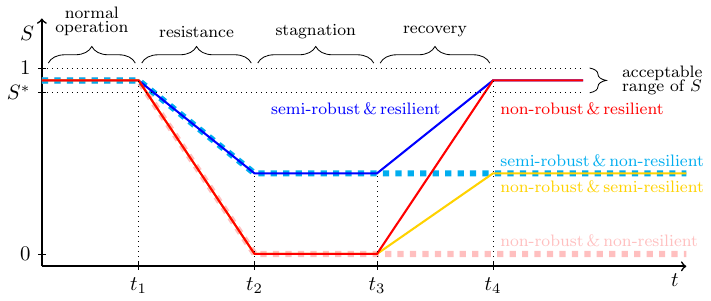}
    \caption{Schematic trapezoidal response trajectory for different types of systems following our terminology.
    The acceptable range of the sustainant $S$ is defined by its lower bound $S^*$.
    The times $t_1, t_2, t_3$, and $t_4$ mark the beginning of the adverse event, the bottoming-out of the sustainant, the beginning of its recovery, and the return to normal operation, respectively.
    As shown, robustness and resilience can be understood as two independent properties. Here, the prefix \emph{semi-} does not necessarily mean \emph{half-} but rather \emph{partially}. The exact quantitative meaning depends on the chosen normalization of each property, as discussed by \citet{ingrisch}.}
    \label{response_trajectory}
\end{figure*}

We definitely agree more with the concept of engineering resilience: If a system can limit the decline of its sustainant, but not recover subsequently, we would not call it resilient. After all, the Latin verb \emph{resilire} from which resilience is derived literally means \emph{to rebound}. We would rather describe such a non-recovering system as \emph{robust}, which aligns with the terminology of \citet{anderies}. Of course, robustness still aids resilience by reducing the magnitude of the necessary recovery. But we argue that robustness and resilience are a sensible separation of system properties, representing the abilities to resist and recover, respectively. In Fig. \ref{response_trajectory}, one can find illustrations of the behavior of systems with different degrees of robustness and resilience.

Recently, the concepts of engineering and ecological resilience have permeated their discipline boundaries: \citet{van2021unifying} define the resilience of an ecological system as its recovery rate. Along with resistance/robustness, they consider resilience to be a constituent of what they call ecological stability (which demonstrates the ambiguity of terminology).

For both engineering and ecological resilience, the established quantification approaches listed by \citet{ingrisch} all revolve around the response quality, like the recovery speed or impact limitation (the latter of which we call robustness). Recently, this paradigm has been weakened by \citet{dakos2022ecological} who interpret ecological resilience as the minimum disturbance magnitude that is required to induce state tipping in a multistable dynamical system. This is very much related to basin stability, as it essentially represents the minimum distance from the attractor to its basin's boundary. It is, therefore, also similar to our concept of basin resilience, albeit only the persistence type.

Basin resilience, as explained in Sec. \ref{sec:framework}, extends the basin concept to the response types of adaptation and transformation, making it applicable to a wider set of systems and models. Meanwhile, it is still flexible regarding the response outcome required for resilience: Recovery speed can be incorporated, as well as impact limitation (in which case we would speak of basin robustness). Essentially, our framework provides a rigorous quantification approach, without being bound to our own understanding of resilience.

\subsection{Resilience Measurement Framework}
\label{sec:framework}

Our proposed resilience measure revolves around the so-called \emph{basin of resilience} which we denote $B_R$. In order to define it, we need to consider the space of adverse influences $\vec{i}$ that can impact the system as well as the space of its possible response options $\vec{r}$. Depending on model complexity, these vector spaces can be of arbitrary dimension. It is important to stress that neither of these spaces are necessarily related to the phase space of the system (which describes its micro-state).
Especially the adverse influence $\vec{i}$ must not be understood as mere displacements of the system's state.

Next, one must define the system's sustainant and be able to evaluate how it reacts over time to each adverse influence $\vec{i}$ in combination with each response option $\vec{r}$. The sustainant might be derived from the system's phase space, or a subspace thereof. But, crucially, it is generally not necessary to exactly know how the micro-state of the system evolves. This means that our framework is not limited to dynamical systems governed by differential equations, but can also be applied to time-discrete and stochastic systems.

Because of this, we treat the sustainant as a function of time, adverse influence and response options taken and, thus, denote it $S(t|\vec{i},\vec{r})$. When $\vec{i},\vec{r}$ are fixed and no confusion is likely to arise, we also use $S(t)$ as a shorthand.

$S(t)$ is typically a bounded quantity. We recommend to normalize $S(t)$ onto the real interval $[0,1]$, with $S(t)=0$ representing complete standstill and $S(t)=1$ representing perfect operation. The sustainant's \emph{acceptable range} is then bounded from above by $S(t)=1$ and from below by a value $S^*$. \citet{ingrisch} refer to this as baseline-normalization.

A useful quantity to define is the deviation of $S(t)$ from its acceptable range, which we call the \emph{sustainant deficit} $\Delta S(t)$:
\begin{equation}
    \Delta S(t) = \max\{0, S^*-S(t)\}.
\end{equation}

The time evolution of the sustainant following an adverse influence $\vec i$ and some specific response $\vec r$ is commonly used \citep{ingrisch, jufri, bie} to illustrate the quality of a system's response. \citet{ingrisch} refer to this time evolution as the \emph{response trajectory} of the system. Typically, the response trajectory of a resilient system will follow a simple pattern: before the adverse event ($t<t_1$), the sustainant rests within the acceptable range of $S^* \leq S(t) \leq 1$, then temporarily falls down ($t_1 < t < t_2$), and eventually rises up, either re-entering the acceptable range or approaching it asymptotically ($t_3 < t < t_4$).
Sometimes, the dip in the response trajectory may have a roughly triangular shape (immediate recovery, $t_2=t_3$), and sometimes a trapezoidal one (delayed recovery, $t_2<t_3$). The latter (in our opinion more general) case is illustrated in Fig. \ref{response_trajectory}. Of course, the exact shape of the trajectory might in reality be less linear, but, in principle, one can always identify the phases of resistance, (optional) stagnation, and recovery.
Without loss of generality, we assume that the system is subject to only a single adverse influence, either temporary or permanent, unfolding at $t=t_1$ and reaching its full impact at $t=t_2$. 

Whether or not a specific response trajectory $S(\cdot|\vec i,\vec r):t\mapsto S(t|\vec i,\vec r)$ exhibits resilient behavior is assessed by a function $\alpha$ which we call the \emph{resilience assessment function}. In places where $\vec i$ and $\vec r$ are clear from the context, we will simply use $S$ as an abbreviation of $S(\cdot|\vec i,\vec r)$. $\alpha$ generally imposes an arbitrary number of conditions $C_1,\dots,C_K$ on the response trajectory $S(\cdot|\vec i,\vec r)$ and checks whether all of them are fulfilled simultaneously.
In the case of a specific individual system with no uncertainties, $\alpha$ is a binary indicator function, taking a value of either 0 or 1. However, generally speaking, the system of interest might have internal uncertainties, or there is a family of similar systems that are examined. Considering this, the resilience assessment function $\alpha$ is defined as the \emph{probability} of the response trajectory fulfilling all conditions:
\begin{equation}
    \alpha(S(\cdot|\vec i,\vec r)) = \text{Pr} \big( C_1(S(\cdot|\vec i,\vec r)) \land \ldots \land C_K(S(\cdot|\vec i,\vec r)) \big),
\end{equation}
where $C_1, \dots, C_K$ are the chosen conditions for the response trajectory $S(\cdot|\vec i,\vec r)$.
Later in this section, some exemplary conditions $C$ will be presented. Note that in order to define $\alpha$, one chooses $C_1,\dots,C_K$ even though we do not denote this dependency explicitly in our notation since that would lead to very clumsy formulas further down.

The basin of resilience $B_R(\vec{r}|\alpha)$, which depends on the chosen response option $\vec{r}$, is now defined as the support of $\alpha$ with respect to adverse influences $\vec{i}$. In words, this means that $B_R(\vec{r}|\alpha)$ is the set of all adverse influences against which the response option has a non-zero chance of providing resilience:
\begin{equation}
    B_R(\vec{r}|\alpha) = \big\{ \vec{i} | \alpha \big( S(\cdot|\vec i,\vec r) \big) > 0  \big\}.
\end{equation}

Our resilience measure, denoted $R$, now reflects the volume of $B_R(\vec{r}|\alpha)$ for each given $\vec r$. However, as the last ingredient of our framework, a density function $\varrho$ is needed to restrict the extent of possible adverse influences $\vec i$ and/or weight them by assigning them different probabilities of occurrence.

The resilience measure $R$ is then defined as the following integral:
\begin{equation}\label{resilience_measure}
    R(\vec{r}|\alpha,\rho) = \int \alpha \big( S (\cdot|\vec{i},\vec{r}) \big) \cdot \varrho(\vec{i}) \,\, d\vec{i}.
\end{equation}

If $\varrho$ is chosen to be an indicator function, $R(\vec r|\alpha,\rho)$ is a kind of fuzzy volume of $B_R(\vec r|\alpha) \cap \text{supp}(\varrho)$ (where $\alpha$ determines the degree of membership). If $\varrho$ is a probability density function, $R(\vec r|\alpha,\rho)$ becomes a probability. Whenever $\alpha$ and $\rho$ are clear from the context, we will abbreviate $R(\vec{r}|\alpha,\rho)$ by $R(\vec{r})$.\\

\noindent For the measure to be meaningful, it is crucial above all to choose sensible definitions of $S$ and $\alpha$.
Deciding on a definition for $S$ may be the hardest part, depending on the individual model's complexity. Regarding $\alpha$, however, we think that some basic resilience conditions $C$ will be universally useful. These exemplary conditions will be presented in the following paragraphs.

In principle, any quantitative measure of the response trajectory can be incorporated into our framework by formulating it as a binary condition $C$, for example using a threshold value. 
However, the following examples will focus on transferring analogous concepts from state-space-based probabilistic stability measures such as basin stability to our sustainant-based framework.

First, we argue that the single key property of a resilient system is that it is able to recover completely, meaning it can permanently restore its sustainant back into the acceptable range, regardless of whether quickly, slowly, or even just asymptotically. This corresponds to the single condition at the heart of the original notion of basin stability \citep{menck}, namely that the system will evolve back into its equilibrium state. We, thus, define our \emph{default resilience condition} $C_0$ using the sustainant deficit $\Delta S$:
\begin{equation}
    C_0(S): \quad \lim_{t \to \infty} \Delta S(t) = 0.
\end{equation}

However, depending on the context, asymptotic recovery may not be sufficient to capture what it means for the system to be resilient. Therefore, an additional condition might be added. This condition is inspired by a variant of basin stability called \emph{finite-time basin stability} which was introduced by \citet{schultz2018bounding}\,. Finite-time basin stability additionally evaluates whether a perturbed system returns into a given neighborhood of an attractor within a given finite time. Translating this to our resilience framework could result in a condition that limits the sustainant deficit $\Delta S$ from a certain point in time onward:
\begin{equation}
    C_F(S|t_F, S_F): \quad \forall t \geq t_F: S(t) \geq S_F,
    \label{c_finite}
\end{equation}
where $t_F$ is the threshold time and $S_F$ is the threshold sustainant value defining the acceptable neighborhood of $S^*$. If a non-asymptotic, finite-time \emph{complete} recovery is desired, the sustainant threshold can be chosen as $S_F = S^*$.

Another more general variant of basin stability is \emph{constrained basin stability} proposed by \citet{van2016constrained}: 
It assesses the intersection of the standard basin of stability with any number of additional basins corresponding to further conditions on the systems transient curve, such as monotonicity in terms of evolution toward the attractor. This variant of basin stability is actually most comparable to our own framework because of the possibility of multiple conditions.

One additional condition that might be relevant when assessing resilience is limiting the area of the trapezoid above the response trajectory, meaning the time integral of the sustainant deficit $\Delta S$. \citet{ingrisch} refer to this quantity as the \emph{perturbation} which we suspect to be potentially confusing; we instead prefer to use the term \emph{cost}. Actually, we define the cost in a slightly more general way:

It may be the case that the perceived damage to the system is not proportional to $\Delta S$. Therefore, we incorporate a weight function $W$ which transforms the sustainant deficit $\Delta S(t)$ into some kind of cost rate. Neither does $W$ need to be bounded, nor does it have to be of any specific dimension (for example currency), but it must be defined such that $W(\Delta S(t))>0$ whenever $\Delta S(t)>0$. Integrating this cost rate along the response trajectory then yields the cost.

The last ingredient required for the aforementioned condition is now a limit $L$ for this cost: 
\begin{equation}
    C_C(S|W,L): \quad \int W\big( \Delta S(t) \big)\, dt  < L.
\end{equation}

A measure related to basin stability, formulated by \citet{hellmann2016survivability}, is \emph{survivability}, which requires the whole phase space transient of the perturbed system to stay within a given desirable set of states. A corresponding resilience condition for the response trajectory could be similar to \eqref{c_finite}, but limiting the sustainant deficit for all times:
\begin{equation}
    C_R(S|S_R): \quad \forall t: S(t) \geq S_R,
\end{equation}
where $S_R$ denotes the threshold sustainant value. The index $R$ is chosen here because this condition actually describes what we interpret as (partial) robustness.\\

Lastly, there is a generalization of basin stability by \citet{mitra2015integrative} called \emph{integral stability} which weights each point in the attraction basin by its largest Lyapunov exponent (a measure of recovery speed). In principle, such a generalization can also be applied to our framework by inserting an appropriate distribution $f(\vec{i})$ into the integral in equation \eqref{resilience_measure}.

\section{Application I: single dynamical node}
\label{sec:application_node}

As a first, easy-to-follow example, we apply our framework to the commonly used dynamical model of a single generator node connected to a large alternating current (AC) grid (see, e.g., \citet{menck2014dead}). We demonstrate that the generalized definition of our resilience framework actually contains the definition of basin stability and its variants. After all, we argue that stability can be interpreted as persistence resilience.

\subsection{System Dynamics}
\label{sec:single_node_dyn}

The single-node system is fully described by the two-dimensional phase space of an oscillator: the voltage phase angle $\theta$ and its oscillation frequency $\omega$. These variables are governed by two first-order differential equations:
\begin{gather}
    \dot{\theta} = \omega\\
    \dot{\omega} = -\delta \times \omega + P - K \times \sin(\theta - \theta_{grid}),
\end{gather}
where $\delta$ is a damping constant, $P$ is the constant net power input to the generator, $K$ is the capacity of the transmission line from the generator to the grid, and $\theta$ is the phase angle of the grid oscillation. 
In our example, we use $\alpha=0.1, P=1, K=8$, and $\theta_{grid}=0$.

The system has an attractor at $(\theta_s, \omega_s) = (\arcsin(P/K), 0)$, as well as an attracting limit cycle at $\omega > 0$.

\subsection{Adverse Influence}

As usual for the single-node system, we consider a displacement of the system in both of its phase space variables. As we explained earlier, not every system is easily modeled as dynamical and allows for phase space considerations, which is why our framework is not limited to this type of adverse influence.

\citet{menck2014dead} show that such a displacement of the node state can be, e.g., the result of shutting down the transmission line between the generator and the grid for some amount of time. Here, for simplicity, we model the displacement as instantaneous, meaning $t_1=t_2$. We define the influence density function $\varrho(\vec{i})$ as a simple indicator function:
\begin{equation}
    \varrho((\theta_0, \omega_0)) = \left\{
    \begin{array}{l}
        1 \quad \text{if }\, \theta_0 \in (-\pi, \pi]  \\
          \qquad \land\; \omega_0 \in [-10, 10]\\[5pt]
        0 \quad \text{else,}\\
    \end{array}
\right.
\end{equation}
where $\theta_0$ and $\omega_0$ are shorthand for $\theta(t_2)$ and $\omega(t_2)$, respectively.

\subsection{Response Options}

In terms of response options, we do not assume any other dynamics than the differential equations described in Sec. \ref{sec:single_node_dyn}. This is what is commonly known as (basin) stability, and we argue that it can be sensibly categorized as persistence resilience: all parameters of the system stay constant after the displacement, yet the system recovers.

\citet{menck2014dead} demonstrate that the parameter $K$ has an impact on the system's basin stability, which means one could interpret $K$ as a response parameter in our framework. To keep this example short, we refrain from varying $K$ and instead calculate the single resilience value for the case $K=8$.

\subsection{Sustainant and Resilience Assessment}

As usual for this system, we treat the attractor $(\theta_s, \omega_s)$ as the desired state of the system, meaning the sustainant $S$ must fulfill $S(\theta_s, \omega_s)=1$. We decide to define $S$ as cosine-like in $\theta$ and Gaussian in $\omega$, with its peak at the attractor:
\begin{equation}\label{eq: single_node_sus}
    S(t) = \frac12 \times (1+\cos(\theta(t)-\theta_s)) \times \exp(-\frac{1}{20}\omega(t)^2)
\end{equation}

The lower bound of the acceptable range is chosen as $S^* = 0.99$\,. To assess the resilience, we evaluate the default cost-limiting condition $C_C(\cdot|W,L)$. The weight function $W(\Delta S)$ is chosen to be the identity function $\mathbb{1}$, and the cost limit is set at $L=12$.

Overall, the formula for our resilience measure becomes
\begin{equation} \label{eq:single_node_res}
    R = \int_{-10}^{10} \int_{-\pi}^\pi \Pr\left(\int_{t_1}^{\infty}\Delta S(t|\theta_0, \omega_0) dt<12\right) d\theta_0 d\omega_0.
\end{equation}

\newpage
\section{Application II: Probabilistic Power Grid}
\label{sec:application_grid}

As our second demonstration, the resilience measurement framework is applied to a time-discrete probabilistic power grid model. It is larger in scale and more complex than the first application, but it is not based on differential equations which makes traditional stability analysis inapplicable.
We simulate a scenario in which the power grid has to adapt to the emergence of prosumers generating their own electricity using photovoltaics (PV).
The prosumers decentralize the energy supply which leads to a re-organization of the power flows.

The goal is to analyze which response options are the most effective for dealing with a wide range of possible prosumer scenarios, as well as how they can be made most efficient.
For simplicity, we choose to assess resilience here solely based on the performance of the power grid after the adverse influence and adaptations have taken place, meaning at $t\geq t_4$.

In the following, the details of the power grid model are kept at the minimum level necessary for understanding the application of the framework. For an in-depth description of all parameters and mechanisms, see the Appendix.

\subsection{Network Structure}

The network structure of power grids is diverse, but mainly determined by their historical development as well as their spatial embedding.
To account for these characteristics, we generate an ensemble of synthetic power grids with a size of $N=100$ nodes using a random growth model provided by the Julia package \emph{SyntheticNetworks} which is based on an algorithm by \citet{synth_net}.

Real-world power grids are predominantly based on alternating current (AC). Accurately modeling AC power flows on networks involves nonlinear equations and synchronization issues in the case of multiple generators. To avoid these, we use the direct current (DC) load flow formulation, which is not necessarily unrealistic as it can also be derived as an approximation of AC flows. It can also be interpreted as a model for future DC-based distribution networks.
This means that we model the power flows with linear equations equivalent to Kirchhoff's laws. The input for these equations is the so-called power injections at each node, which are the difference between the nodes' respective demand and supply, and the outputs are the power flows on each line.

To balance any mismatch between the grid's total demand and supply, one node is chosen as the so-called slack bus. Since this role is typically associated with a large power plant or external grids, the role of the slack bus is assigned permanently to the node with maximal closeness centrality for each generated graph. The remaining 99 nodes start off as consumer nodes and are later partially converted to prosumer nodes. 
As a consequence, in the absence of prosumers, the slack bus is the sole provider of energy to the grid.

\subsection{Power Injections}
\label{sec:power_injections}

The power demand and supply at each node are modeled stochastically by drawing values from two separate data sets.
These data sets are manually selected subsets of real-world time series of household power consumption and PV generation, respectively.
Both time series data sets were modified to have a matching temporal resolution of $\Delta t=60\,\text{s}$ and divided into daily chunks, starting and ending at midnight. From each data set, 24 daily chunks were selected, covering the full cycle of the year relatively evenly with about two days per month and, thus, capturing the seasonal variability.
When simulating a longer period of time, these daily chunks are randomly chained together to generate unique time series for every node. This generation process means that the time of day is always aligned for all nodes, but the time of year is not consistent. This makes the model less realistic, but maximizes the time series variability achievable with the relatively small data sets.

Both power data sets are separately normalized to a mean value of 1 arbitrary power unit (p.u.). In the model, all consumers and prosumers have identical average demand of $\overbar{d} = 1$\,p.u., and all prosumers generate an identical average amount of PV power supply $\overbar{s}$, which is scaled up or down depending on the exact scenario.
    

\subsection{Adverse Influence}

The adverse influence $\vec{i}$ of the power grid system is the unregulated conversion of random consumers into prosumers. Since only the final performance of the power grid will be assessed, the order and speed of the prosumers' appearance is irrelevant. As mentioned in Sec. \ref{sec:power_injections}, a major simplification in the model is that all prosumers have an identical average power supply.

To create a space of many possible adverse influences, they are modeled to be dependent on two parameters: First, the number of consumers turning into prosumers is denoted $n_p$. Second, the ratio of the prosumers' PV power production to their power consumption is controlled by the parameter $r_p$, which is used as a scaling factor for the PV generation time series. The range of possible values for $n_p \in [1,99]$ is straightforward. For the second parameter, the interval is chosen to be $r_p \in [0.1, 10]$, which covers one order of magnitude for both under- and overproduction.

A simple probability density function $\varrho$ is constructed by assuming a linear decrease in probability for both parameters:
\begin{equation}
\begin{gathered}
    \varrho(\vec{i}) = \varrho((n_p, r_p)) = k \times (100-n_p) \times (10-r_p)\\
    \text{with $k \approx 4.12244 \times 10^{-6}$}\\
    \text{such that } \sum_{n_p=1}^{99} \int_{0.1}^{10} \varrho((n_p, r_p))\, dr_p = 1 .
\end{gathered}
\label{k}
\end{equation}

\begin{figure}
    \centering
    \adjincludegraphics[width=\linewidth,trim={0 0 {.075\width} 0},clip]{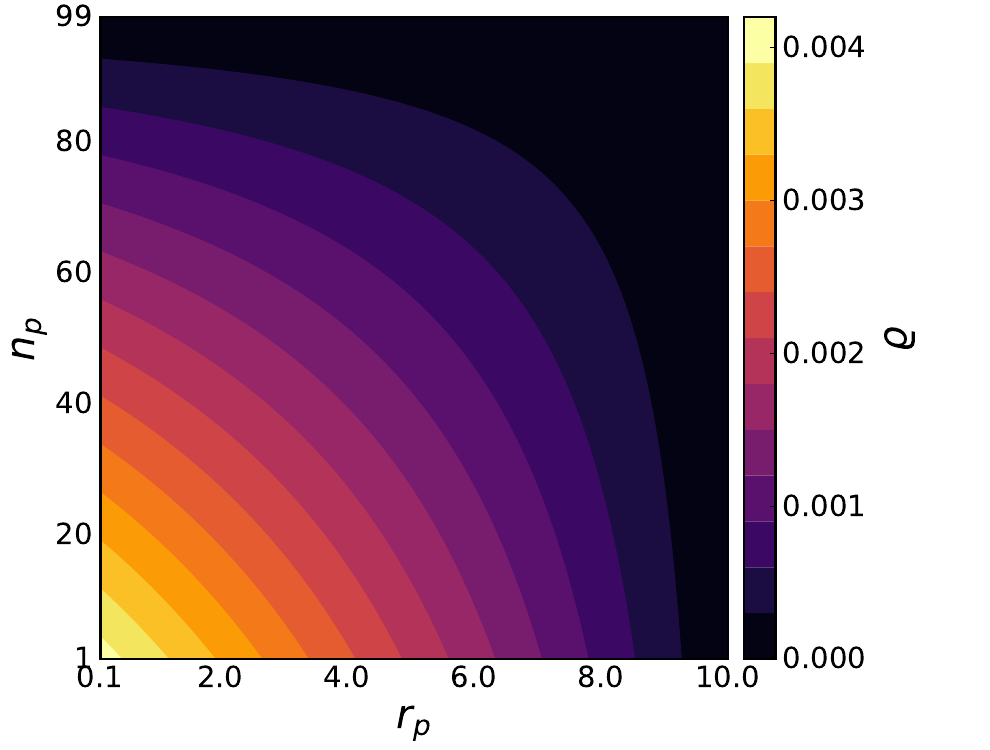}
    \caption{Visualization of the adverse influence probability density $\varrho((r_p, n_p)).$}
    \label{density_plot}
\end{figure}

An illustration of the density function $\varrho$ can be found in Fig. \ref{density_plot}.

\subsection{Vulnerabilities}

The vulnerability of the power grid consists in the power flow capacities of the power lines. If they are insufficient for the given power injections, the performance of the grid suffers, which will be captured by the sustainant.

\begin{figure}
    \centering
    \adjincludegraphics[width=\linewidth,trim={{.075\width} {.075\width} {.075\width} {.075\width}},clip]{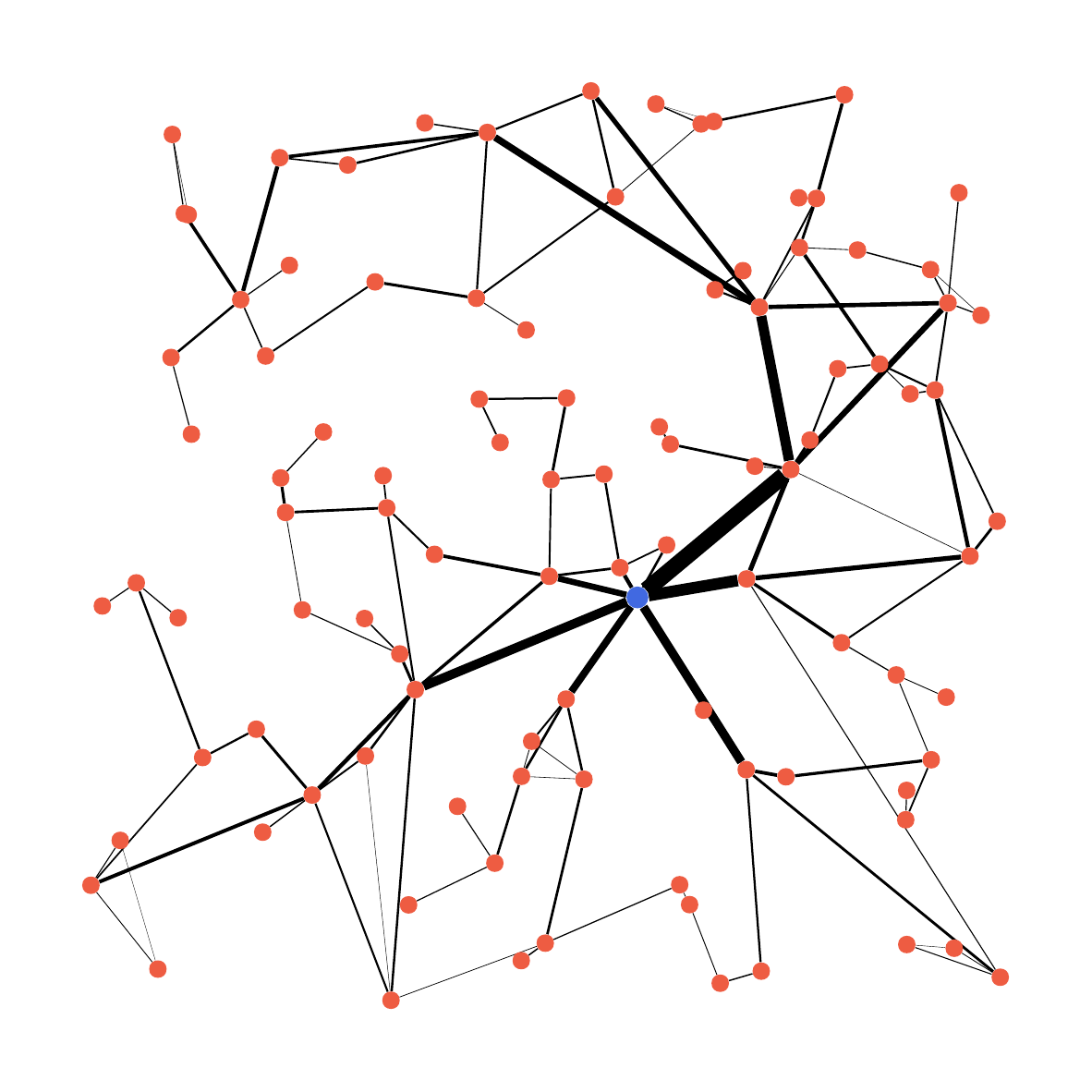}
    \caption{Example of a randomly generated power grid graph in its initial state (without prosumers). The slack bus is highlighted in blue, all other red nodes are consumers. The edge widths proportionally represent the initial capacities of the power lines, which are optimized for centralized supply by the slack bus.}
    \label{grid_example}
\end{figure}

After generating each random grid, the initial line capacities are set up in such a way that the grid is optimized for the traditional, centralized power supply by the slack bus.
This is achieved by generating random power demand time series where all nodes are treated as regular consumers, and assessing the maximum flows that occur.
An example of the resulting initial capacity distribution is illustrated in Fig. \ref{grid_example}.

Whenever a power flow calculated from the injections exceeds the capacity of the corresponding power line, the line shuts down, which means the edge is removed from the graph. This necessitates a redistribution of power flows and may cause a cascade of further line failures.

The cascades are assumed to happen so quickly that the power injections effectively do not change while they unfold. It is also assumed that the overloaded lines reboot at a similarly short time scale. This separation of time scales means that the damage caused by the line failures is confined to the single simulation time step in which they are triggered. Thus, for every time step, the grid is modeled to start off in its completely intact state.

Since the time step size is dictated by the power data time series, the duration of line shutdowns must be $\Delta t\leq60\,\text{s}$, which we estimate as $\Delta t=60\,\text{s}$. Notably, this actually matches the maximum \emph{fault-ride-through} duration for power line faults defined in the regulations \citep{reboot} of the German Association of Electrical Engineering (VDE).

If the cascades result in a disconnected grid, only one of its connected components contains the slack bus. Within this component, the slack bus can adjust the overall injection balance to zero. In the other component(s) however, there will either be a power surplus or a power deficiency. In both cases, the component's balance has to be adjusted to 0 to make a follow-up linear flow calculation possible.

In the first case, all nodes with positive injections (supply $>$ demand)  within the component equally and instantly \emph{waste} their excess power supply using some kind of wasting mechanism (like a high-power radiator or a high-resistance ground wire).
In the second case (demand $>$ supply), a local blackout happens and all nodes within the component have to set their injections to zero. Those nodes with previously positive injections do this by wasting the injected part of their supply. Nodes with previously negative injections instead have an internal \emph{lack} of power supply.
Both wasted power and lacking power are represented by the same variable called the \emph{mismatch} $m$. Wasted power is registered as $m>0$ and lacking power as $m<0$.

After all components' balances are adjusted to 0, the adjusted injections are fed back into the linear flow equation, producing the redistributed flows. This procedure is repeated until a stable flow pattern emerges, meaning there are no more line capacity violations. In the most extreme case, it could mean that \emph{all} lines in the power grid have failed, leaving none to continue the cascade.

\subsection{Response Options}

To cope with the emergence of the prosumers, two separate response options $\vec{r}$ of the power grid are examined. One of them falls into the category of adaptation, and the other one is an example of transformation. Both of these response options are actually families of response strategies, each dependent on two parameters: Power line capacity upgrades are controlled by the line budget $\phi_L$ and the line allocation parameter $\varepsilon$. Battery installation is dependent on the battery budget $\phi_B$ and the battery threshold parameter $\lambda$.

Again, the speed and order of line upgrades or battery installations is irrelevant for the resilience assessment, allowing for a simpler model.\\

\subsubsection{Power Line Capacity Upgrades}

The first, adaptive response strategy consists in upgrading the capacities of the existing power lines, making them less prone to overloading and subsequent shutdown. The main parameter of this response strategy is called the \emph{line budget}, representing the amount of material from which to build the line capacity upgrades. This budget has the dimension of \emph{length $\times$ power}, since the material cost of a power line upgrade depends proportionally on both the capacity difference $\Delta c$ (which has the dimension of power) and the metric length $l$ of the line. For simplicity, additional construction costs are omitted in this model.

The number of edges in each generated power grid, their metric lengths, and the initial capacities are all subject to random fluctuations. It is, therefore, sensible to not use the absolute line budget as a parameter, but instead a proportionality factor that relates the adaptation budget $\beta_{L}$ to the initial cumulative line budget $\beta_{L,0}$.
This factor is denoted $\phi_L$:
\begin{equation}
    \beta_L =  \phi_L \times \beta_{L,0} \qquad \text{with }\beta_{L,0} = \sum_e c_{e,0}\times l_e,
\end{equation}
where $e$ is an edge index, $c_e$ is the edge's flow capacity, and $l_e$ is its Euclidean length.

The line adaptation budget $\beta_L$ is allocated non-uniformly to all existing power lines, depending on the secondary adaptation parameter which is called the \emph{line allocation parameter} $\varepsilon$.
Specifically, every prosumer node contributes to every line's upgrade, and the contribution size depends on the distance between the two via the following power law:
\begin{equation}
    \delta c_{e,n} \propto d^*_{e,n}{}^\varepsilon,
\end{equation}
where $\delta c_{e,n}$ is the contribution of node $n$ to the capacity upgrade of edge $e$, $d^*_{e,n}$ is the shortest path distance between the two, and $\varepsilon$ is the line allocation parameter.

Heuristically, it makes sense to limit the range of the allocation parameter to $\varepsilon \leq 0$ because the line overload risk will be higher closer to a prosumer node. The edge case of $\varepsilon = 0$ results in a uniform capacity upgrade for all lines, and lower values increasingly prioritize lines closer to prosumers.

We hypothesize that the optimal strategy consists in non-uniform upgrades with $\varepsilon<0$ because power flows will tend to split up into smaller flows with increasing shortest path distance to their injection node. The optimal value of $\varepsilon$ is likely independent of the budget $\phi_L$, but the exact value of $\varepsilon$ probably becomes less relevant above a certain budget value $\phi_L$.\\

\subsubsection{Battery Installation}

The second, transformative response strategy is the installation of batteries at prosumer nodes. Their purpose is to reduce peaks and troughs in the power injections of the prosumer nodes, which, in turn, reduces the risk of line overloads.

The strategy's primary parameter is analogous to the primary one of line upgrades: The \emph{battery budget} $\beta_B$ represents the material from which to build the batteries, which is simplified as the cumulative amount of energy that can be stored in all batteries throughout the grid. Again, for generalization purposes, the budget is represented as the multiple of a reference energy value $E_0$. This reference value is chosen to be the average grid-wide energy demand during one day, and the proportionality factor is called $\phi_B$. Since all consumers are modeled to have equal average demand $\overbar{d}$, the budget is allocated uniformly onto all prosumer nodes:
\begin{equation}
\begin{gathered}
    \beta_B = \phi_B \times E_0,\\
    b_{\text{max}} = \frac{1}{n_p} \times \beta_B,
\end{gathered}
\end{equation}
where $b_{\text{max}}$ is the maximal energy content of each battery.

Consequently, the secondary adaptation parameter does not manage the budget allocation. Instead, it controls how strongly the batteries react to peaks and troughs in the power injections.
Since the batteries have finite capacity, they must have a net power output of zero, which means that a prosumer node's mean power injection cannot be altered. Only the deviations from the injection mean can be modified.

The batteries are modeled in a way such that they aim at imposing a maximum and value on the node's power injection. Thus, when the prosumer's power balance exceeds this range, and if the battery has the required energy capacity in that moment, they absorb a part of the prosumer's power balance.
After having installed the batteries at $t=t_4$, they are initiated at 50\,\% charge so that they can absorb injection deviations in both directions equally well.

To control the battery behavior, the \emph{battery threshold parameter} $\lambda$ is introduced. Together with the influence parameters $n_p$ and $r_p$ of the respective scenario, it defines the maximum and minimum injection of prosumers:
At $\lambda = 1$, all deviations from the mean are absorbed, regardless of $n_p$ and $r_p$. 
At $\lambda = 0$, no deviations are absorbed, making the batteries useless. Therefore, the sensible range for the battery threshold parameter is $0< \lambda \leq 1$.
A relevant value in between is $\lambda = \nicefrac{1}{9,900} \approx 10^{-4}$, where all deviations are absorbed only in case of the most mild adverse influence $(n_p=1, r_p =0.1)$. For even lower values of $\lambda$, some deviations are allowed even in that most mild case.

Our intuitive hypothesis is that there exists an optimal threshold value $\lambda <1$ which increases linearly with increasing battery budget $\phi_B$. This optimal value represents the sweet spot between absorbing too little such that the allowed injection fluctuations cause line outages, and absorbing too much such that the battery quickly becomes full or empty and cannot absorb any further fluctuations.

\subsection{Sustainant and Resilience Assessment}

Selecting the sustainant of the system is arguably the most controversial part of the modeling process.
For the prosumer-based power grid, we want to aim at avoiding not only any lack of power supply for regular consumers, but also the waste of excess power that prosumers may have to resort to. After all, with the decentralization of power supply, the power grid's purpose essentially becomes distributing all available power as efficiently as possible.

To reflect this, we define the transmission efficiency $\tau$ based on the sum of absolute power injections $|j|$ fed into the power grid at any given time step, excluding the slack bus. From this sum, we subtract all wasted power and all lacking power, which are both encoded in the mismatch variable $m$ (distinguished by their signs). To treat both equally, we subtract the absolute value $|m|$.
The transmission efficiency $\tau$ is now defined as the remaining fraction of the injection sum which can be realized by the power grid:
\begin{equation}
    \tau(t) = \left\{
    \begin{array}{l}
        \dfrac{\sum_{n \neq n_S} |j_n(t)|-|m_n(t)|}{\sum_{n \neq n_S} |j_n(t)|}\\[12pt]
        \hspace{40pt} \text{if }\, \sum_{n \neq n_S} |j_n(t)| \neq 0\\[10pt]
        1 \hspace{35pt} \text{else,}\\
    \end{array}
\right.
\end{equation}
where $n$ is a node index and $n_S$ is the node index of the slack bus.

Due to the stochastic nature of the power data and the limited amount of time steps to base the initial line capacities on, this efficiency inevitably fluctuates a lot over time. Consequently, using $\tau$ directly as the sustainant $S$ would necessitate an unreasonably low boundary $S^*$ for the acceptable range.
To avoid this, the decision is made to observe only the \emph{average} transmission rate, which is calculated over the span of 60 days:
\begin{equation}
    \begin{gathered}
    \forall t_0 \leq t \leq t_0 +60\,\text{d}:\\[5pt]
    S(t_0) = \overbar{\tau(t_0 \leq t \leq t_0 +60\,\text{d})},
    \label{sustainant}
\end{gathered}
\end{equation}
where $t_0$ is a time step index. 

For this average efficiency, the acceptable range of $S$ can be defined more narrowly. We decide to set the goal of limiting the average sustainant deficit to a value that corresponds to one of the 99 nodes having its injection unfulfilled for one entire day per year:
\begin{equation}
    S^* = 1 - \frac{1}{99 \times 365} \approx 1- 2.7674 \times 10^{-5}.
    \label{s_star}
\end{equation}

As mentioned before, we decide to define the resilience assessment function $\alpha$ based solely on the post-adaptation performance of the power grid ensemble. This corresponds to the finite-time resilience condition $C_F(\cdot|t_F,S_F)$ defined earlier, with $t_F = t_4$ and $S_F = S^*$. This condition also implies the default condition $C_0$ (asymptotic recovery):
\begin{equation}
    \alpha(S) = \text{Pr} \big(\forall t \geq t_4: S(t) \geq S^* \big).
\end{equation}


Given the stochastic nature of the power grid model, we deem it sufficient to observe the sustainant for 60 days after implementing the response. We further calculate the probability $\alpha$ based on a sample of 50 randomly generated power grids.

Combining this with all previous definitions, the formula for the power grid resilience measure becomes
\begin{widetext}
\begin{equation}
    R(\vec{r}|t_4,S^*) = \sum_{n_p=1}^{99} \int_{0.1}^{10}
    \text{Pr} \big(\forall t_4 \leq t \leq t_4 +60\,\text{d}:
    S(t|\vec{r}, (r_p, n_p)) \geq S^* \big)
    \times k \times (100-n_p) \times (10-r_p) \, dr_p,
\end{equation}
\end{widetext}

where $S^*$ and $k$ are two constants defined in \eqref{s_star} and \eqref{k}, respectively.

This integral-sum combination is estimated by quasi-Monte-Carlo (QMC) sampling using 256 samples. The samples are generated from the Sobol sequence \citep{sobol} which was chosen to allow for later addition of further samples. However, since the resolution proved sufficient, there were no samples added.

\section{Results}
\label{sec:results}

\subsection{Single-Node Persistence}

\begin{figure}
    \centering
    \adjincludegraphics[width=\linewidth,trim={{.03\width} {.025\width} {.025\width} {.025\width}},clip]{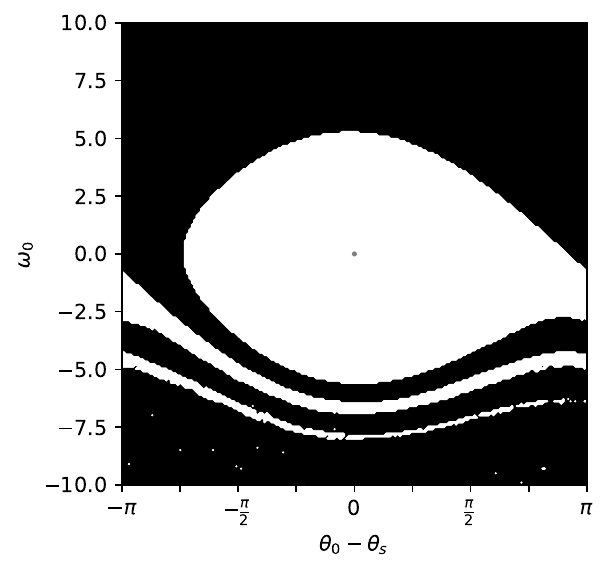}
    \caption{Basin of cost-limiting persistence resilience (or stability) for the single-node system based on equation \eqref{eq:single_node_res}. The central dot marks the attractor state $(\theta_s,0)$, the white region is the resilience basin within the space of displacements $(\theta_0, \omega_0)$.}
    \label{fig:single_node_basin}
\end{figure}

Estimating the resilience value from equation \eqref{eq:single_node_res} with 40,000 samples yields $R\approx 47.31$\,. This value may not be intuitively interpretable without the phase space volume of the sampling region ($V\approx 125.66$) as a reference. But due to the chosen cost-limiting resilience condition, the basin is bounded. This means that its volume, and therefore $R$, is independent of the integration boundaries in $\omega_0$ (given that the latter is sufficiently large). An illustration of the resilience basin can be found in Fig. \ref{fig:single_node_basin}.

\subsection{Power Grid Adaptation}
\begin{figure*}
    \centering
    
    \begin{subfigure}[t]{0.505\textwidth}
        \vskip 0pt
        \centering
        \includegraphics[width=\linewidth]{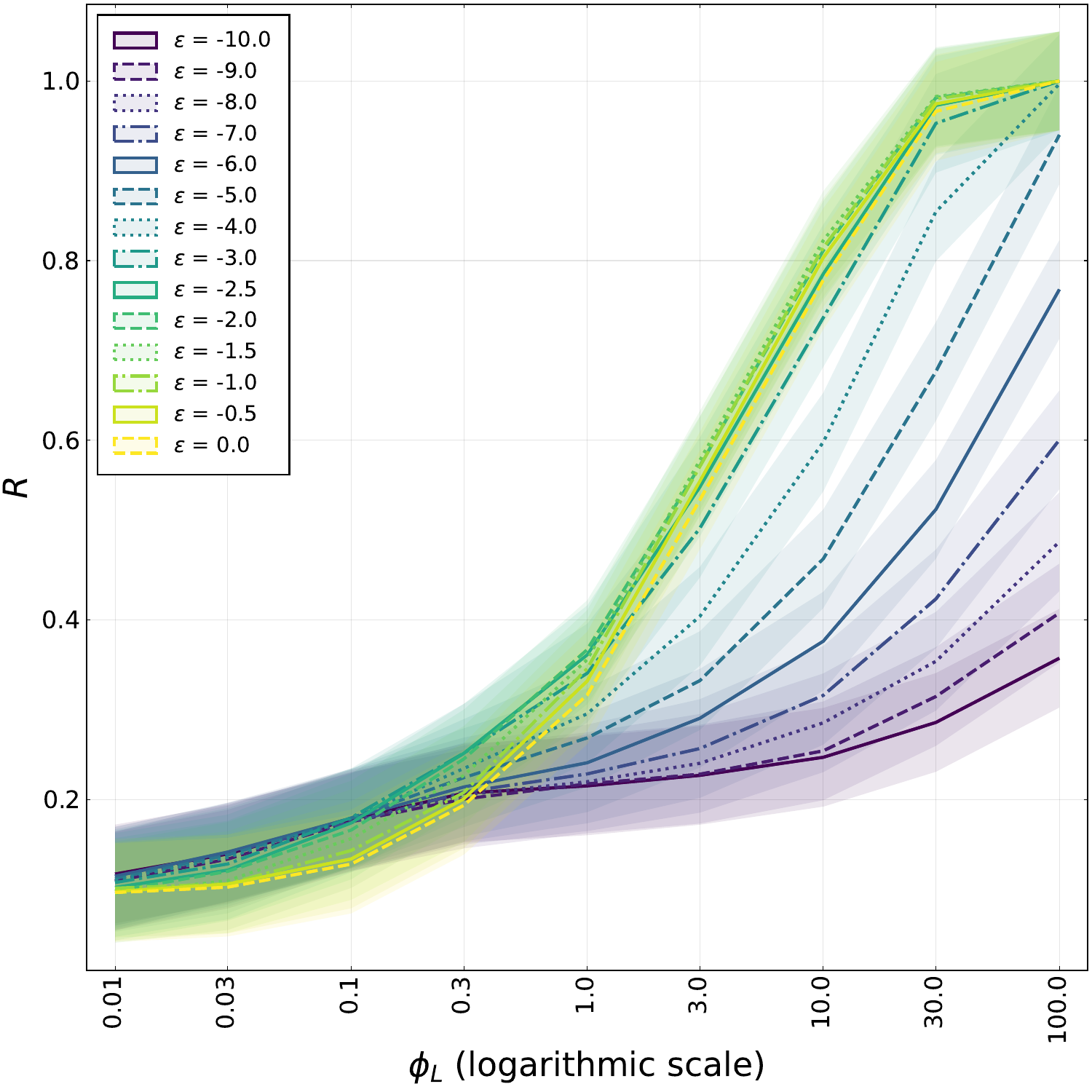}
        \caption{Line budget dependency $R(\phi_L)$ for different values of $\varepsilon$}
        \label{line_budget_plot}
    \end{subfigure}
    \hfill
    \begin{subfigure}[t]{0.485\textwidth}
        \vskip 0pt
        \centering
        \includegraphics[width=\linewidth]{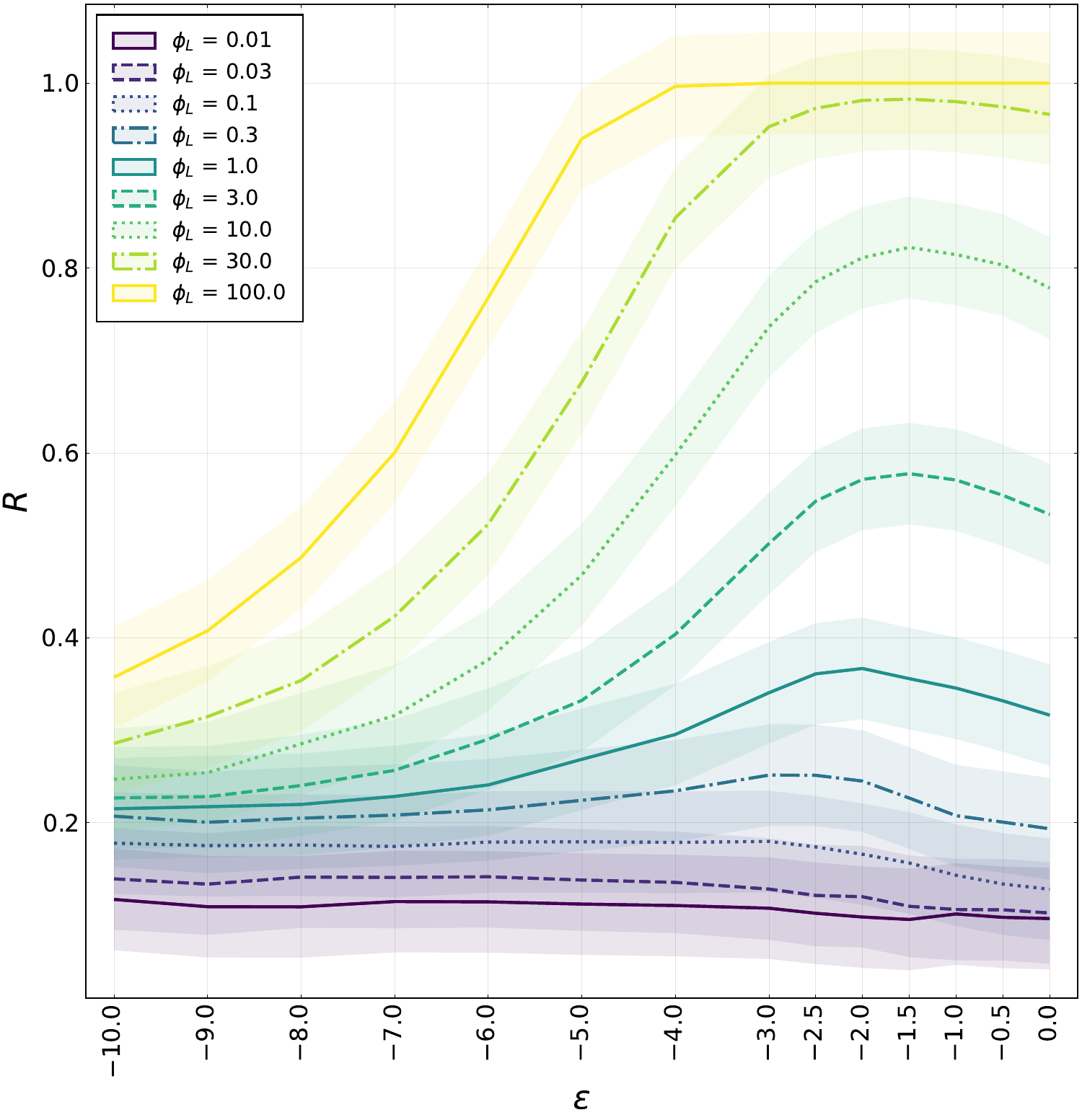}
        \caption{line allocation parameter dependency $R(\varepsilon)$ for different values of $\phi_L$}
        \label{line_secondary_plot}
    \end{subfigure}
    
    \caption{Response parameter dependency of the resilience measure $R(\vec{r})$ for power line capacity upgrades. The actual data points calculated from QMC sampling are indicated by the ticks on the parameter axes, the lines in between are linear interpolations. The transparent ribbons indicate the standard error from Monte Carlo estimation.}
    \label{line_resilience}


    \centering
    
    \begin{subfigure}[t]{0.505\textwidth}
        \vskip 0pt
        \centering
        \includegraphics[width=\linewidth]{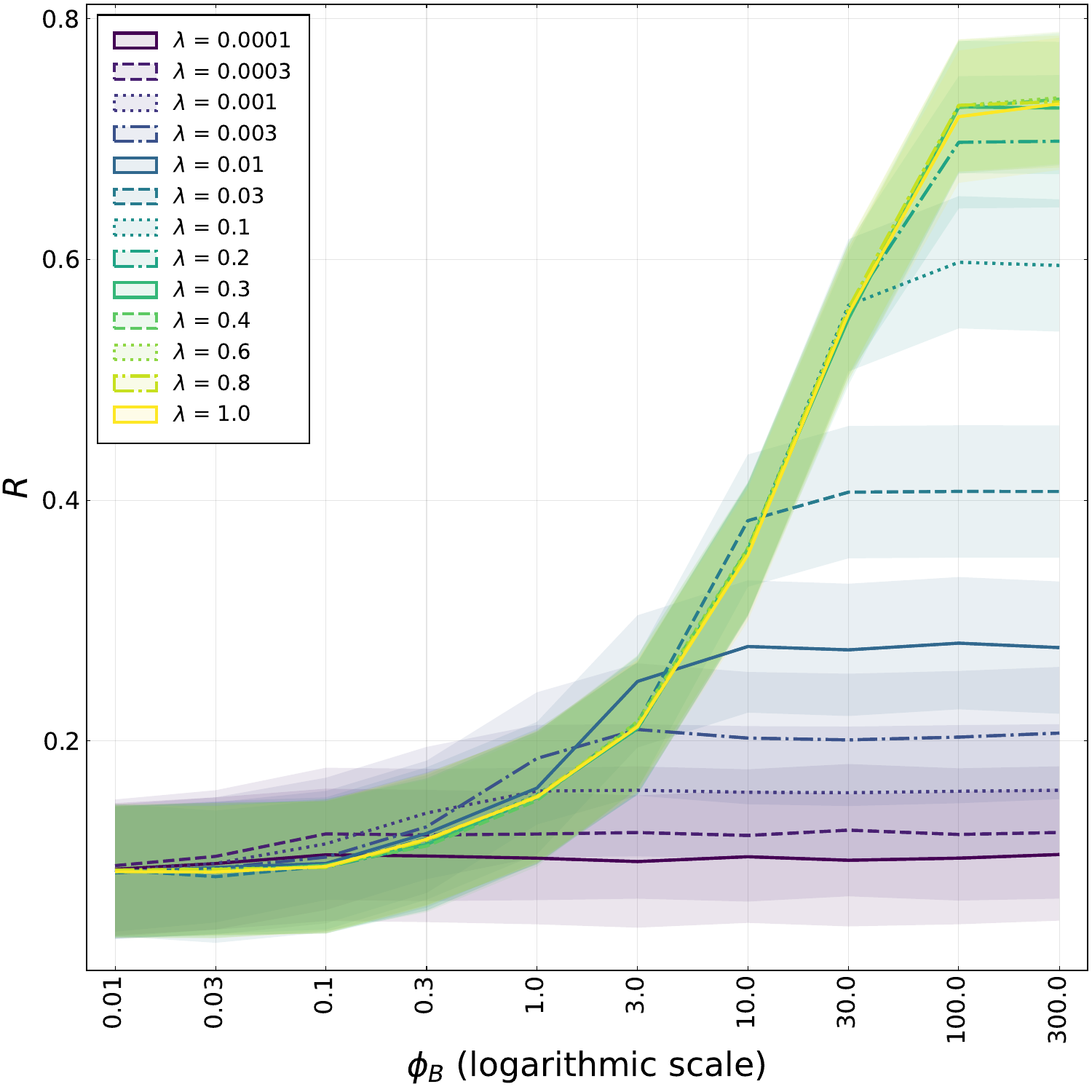}
        \caption{Battery budget dependency $R(\phi_B)$ for different values of $\lambda$}
        \label{battery_budget_plot}
    \end{subfigure}
    \hfill
    \begin{subfigure}[t]{0.485\textwidth}
        \vskip 0pt
        \centering
        \includegraphics[width=\linewidth]{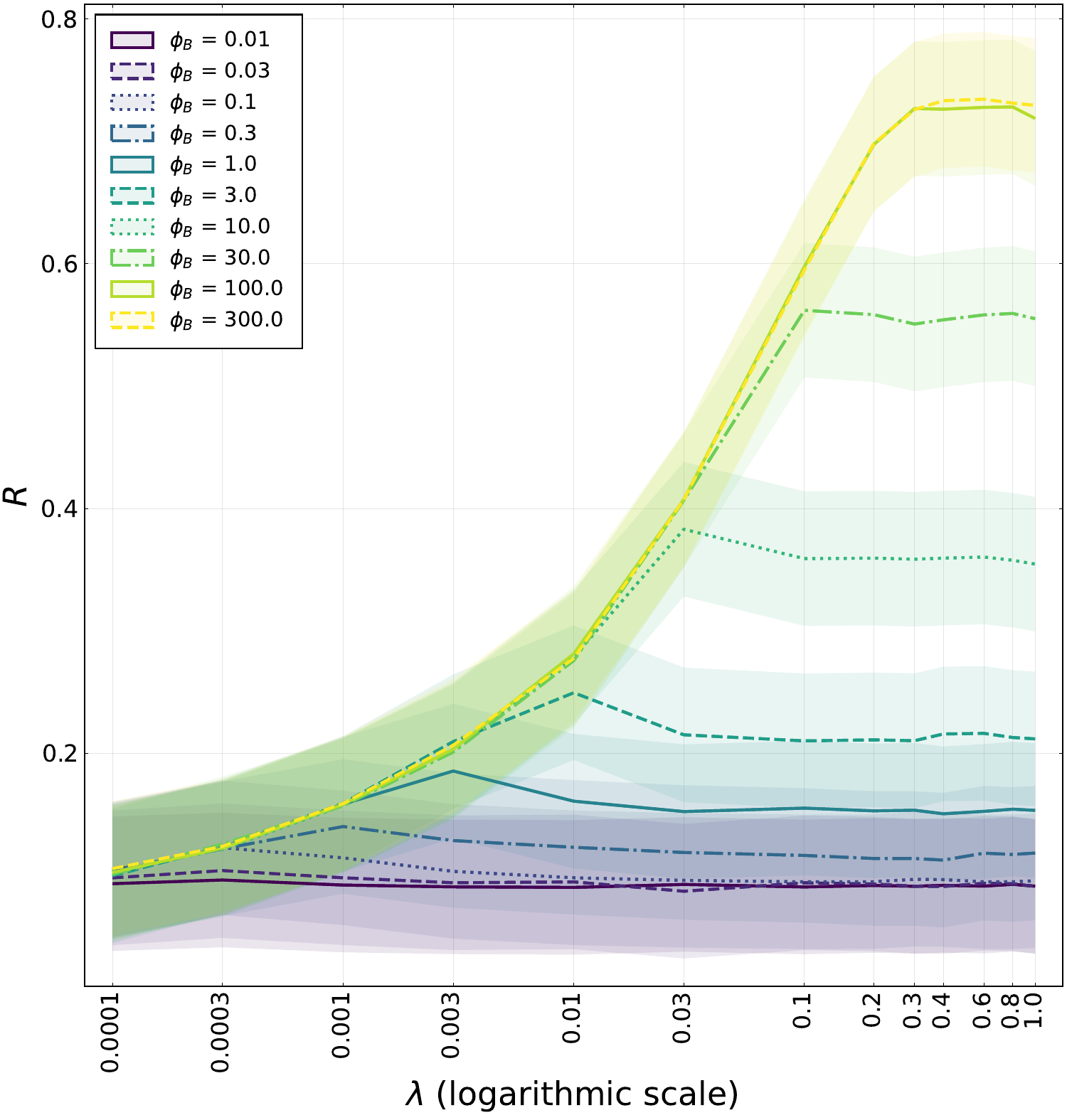}
        \caption{battery threshold parameter dependency $R(\lambda)$ for different values of $\phi_B$}
        \label{battery_secondary_plot}
    \end{subfigure}
    
    \caption{Response parameter dependency of the resilience measure $R(\vec{r})$ for battery installation. The actual data points calculated from QMC sampling are indicated by the ticks on the parameter axes, the lines in between are linear interpolations. The transparent ribbons indicate the standard error from Monte Carlo estimation.}
    \label{battery_resilience}
\end{figure*}

First and most strikingly, the maximum achieved resilience value is different for both response strategies, as can be seen when comparing Figs. \ref{line_resilience} and \ref{battery_resilience}: With line upgrades, the power grid can reach $R=1$, meaning there is a guaranteed recovery from all adverse influences. For battery installation, however, the curve plateaus at a significantly lower value of $R \approx 0.7$.\\

\begin{sidewaysfigure}[p]
    
    \begin{subfigure}[t]{0.188\linewidth}
         \centering
         \includegraphics[width=\textwidth]{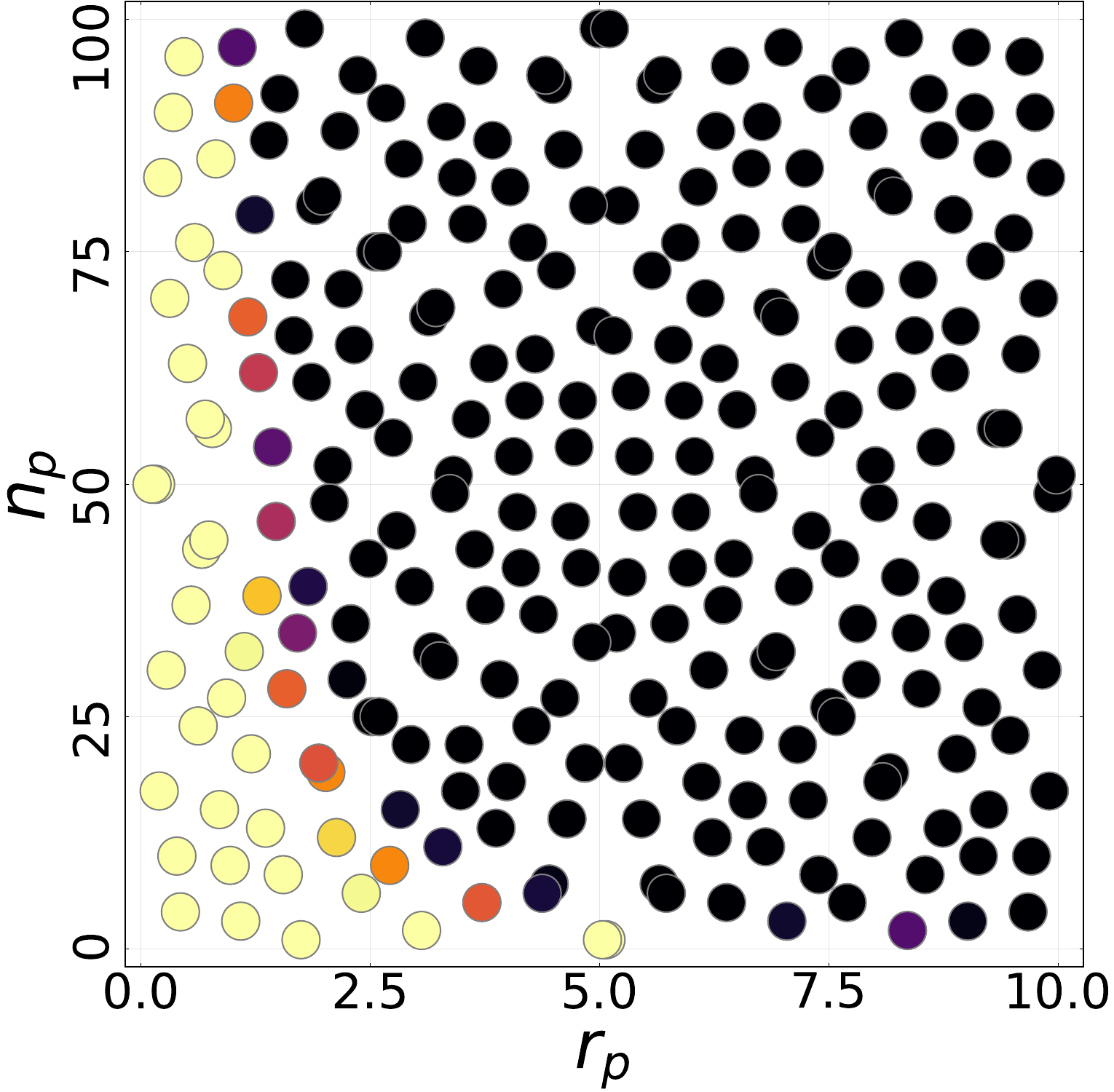}
         \caption{$\phi_L=1$}
     \end{subfigure}
     \hfill
     \begin{subfigure}[t]{0.188\linewidth}
         \centering
         \includegraphics[width=\textwidth]{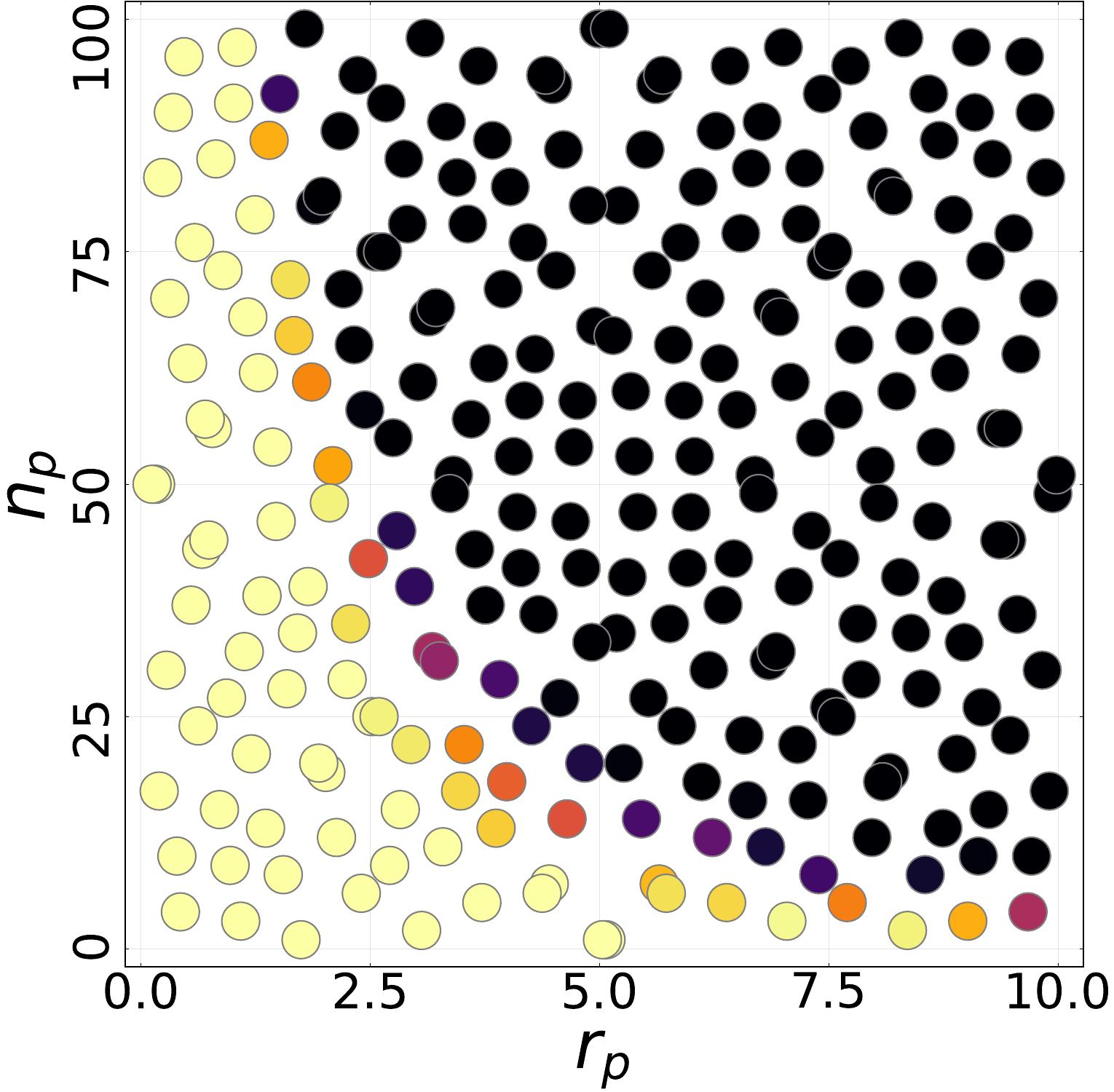}
         \caption{$\phi_L=3$}
         \label{qmc_line_net_neutral}
     \end{subfigure}
     \hfill
    \begin{subfigure}[t]{0.188\linewidth}
         \centering
         \includegraphics[width=\textwidth]{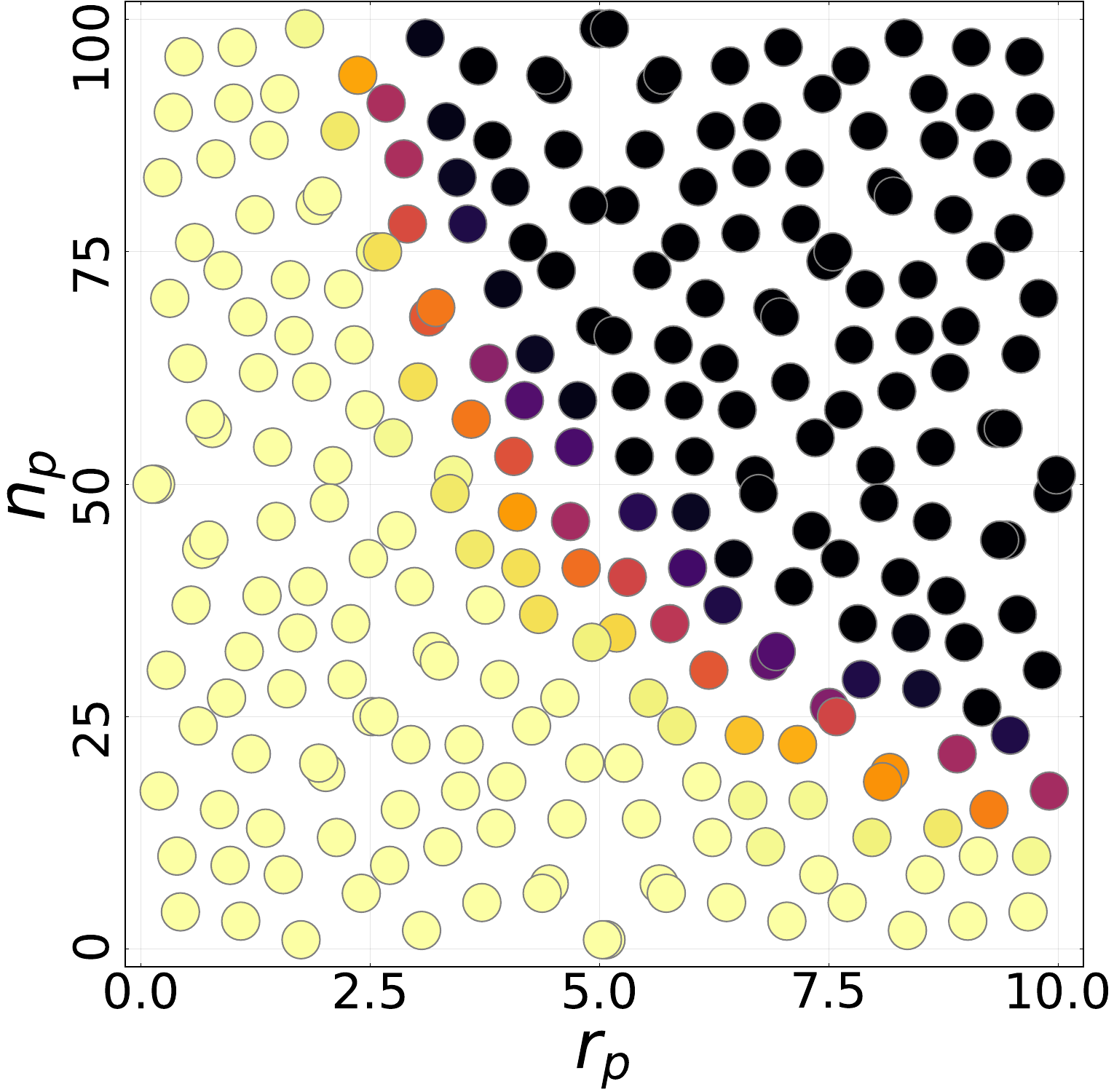}
         \caption{$\phi_L=10$}
     \end{subfigure}
     \hfill
     \begin{subfigure}[t]{0.188\linewidth}
         \centering
         \includegraphics[width=\textwidth]{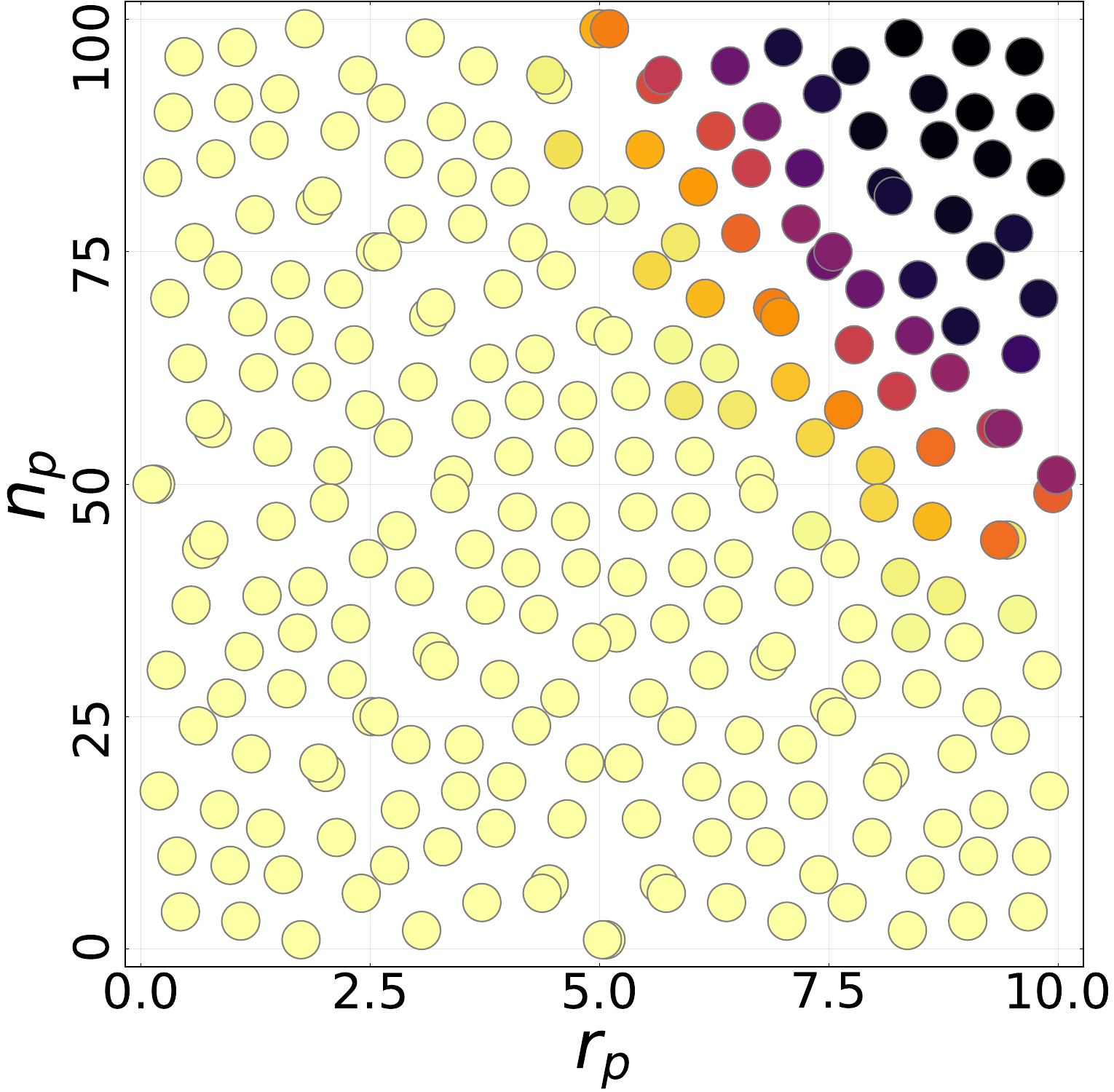}
         \caption{$\phi_L=30$}
     \end{subfigure}
     \hfill
     \begin{subfigure}[t]{0.222\linewidth}
         \centering
         \adjincludegraphics[width=\textwidth,trim={0 0 {.08\width} 0},clip]{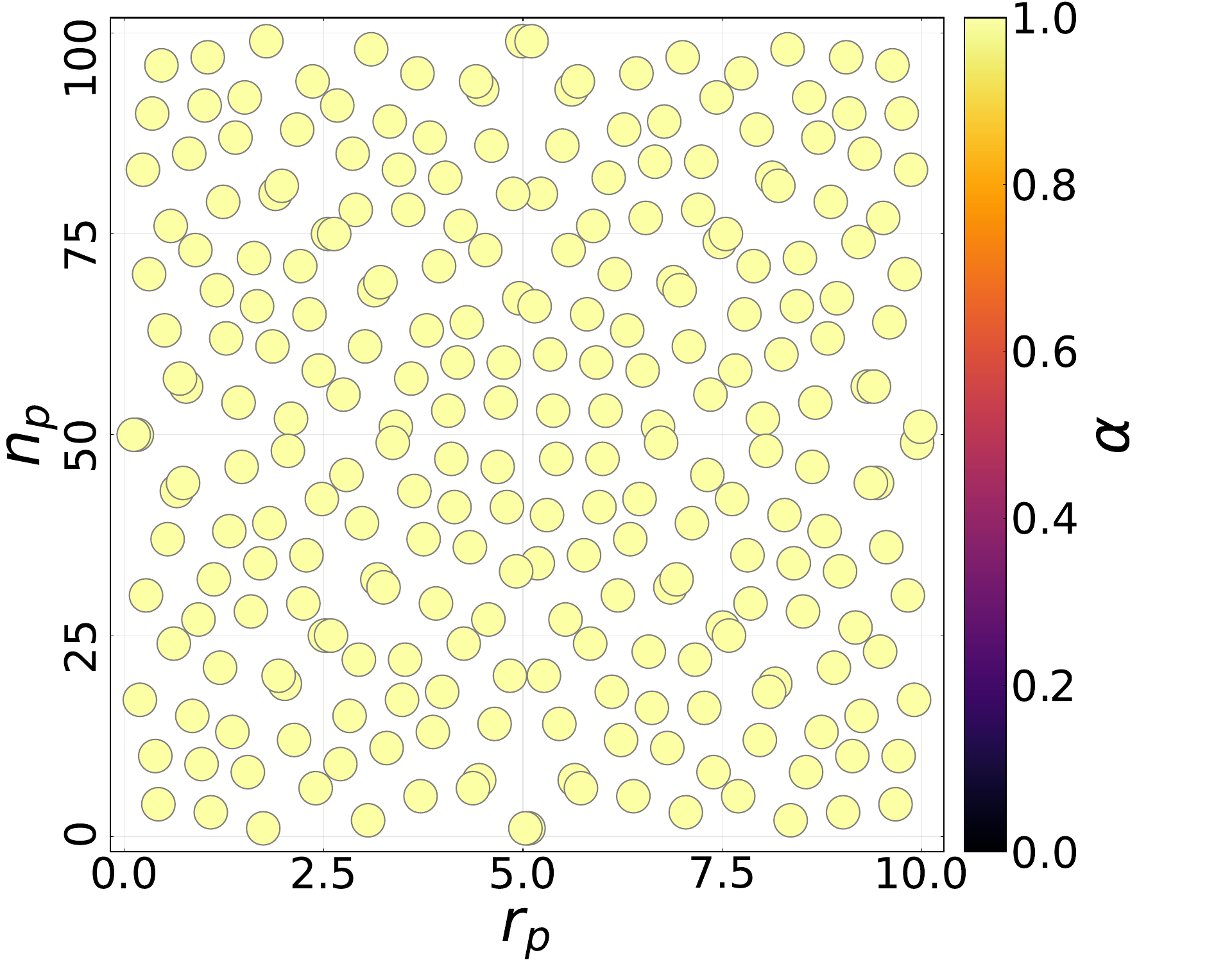}
         \caption{$\phi_L=100$}
     \end{subfigure}
    
    \caption{Resilience basins for line capacity upgrades based on QMC sampling of the resilience assessment function $\alpha(r_p, n_p)$. The line allocation parameter is fixed at $\varepsilon=-1.5$ and the line budget $\phi_L$ is increased incrementally with each subfigure. The vertical axis indicates the number $n_p$ of appearing prosumers and the horizontal axis their ratio $r_p$ of power production to consumption.}
    \label{qmc_line}
    \vspace{0.5cm}
    \end{sidewaysfigure}

\begin{sidewaysfigure}
     \begin{subfigure}[t]{0.188\linewidth}
         \centering
         \includegraphics[width=\textwidth]{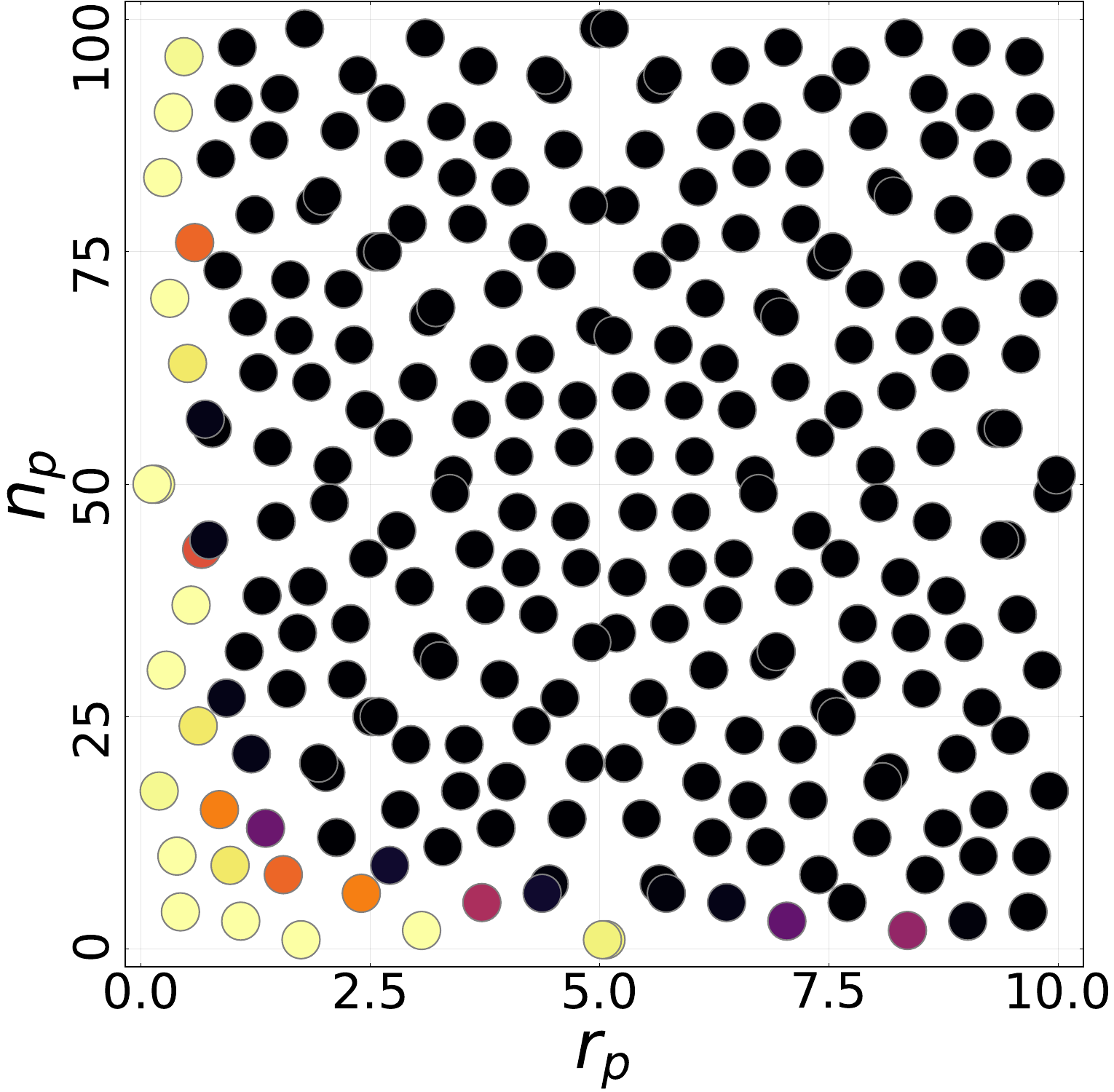}
         \caption{$\phi_B=3$}
     \end{subfigure}
     \hfill
     \begin{subfigure}[t]{0.188\linewidth}
         \centering
         \includegraphics[width=\textwidth]{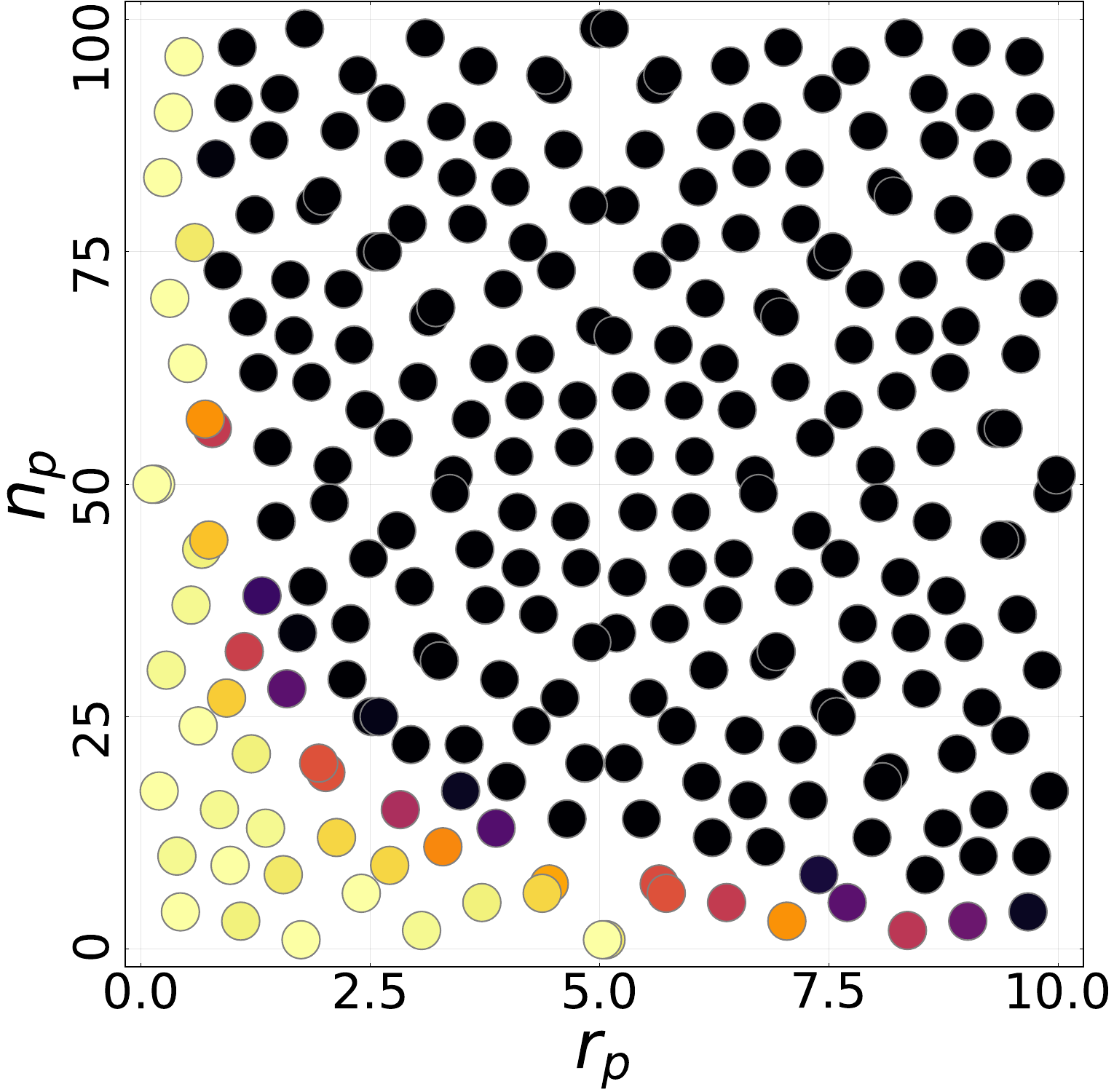}
         \caption{$\phi_B=10$}
     \end{subfigure}
     \hfill
    \begin{subfigure}[t]{0.188\linewidth}
         \centering
         \includegraphics[width=\textwidth]{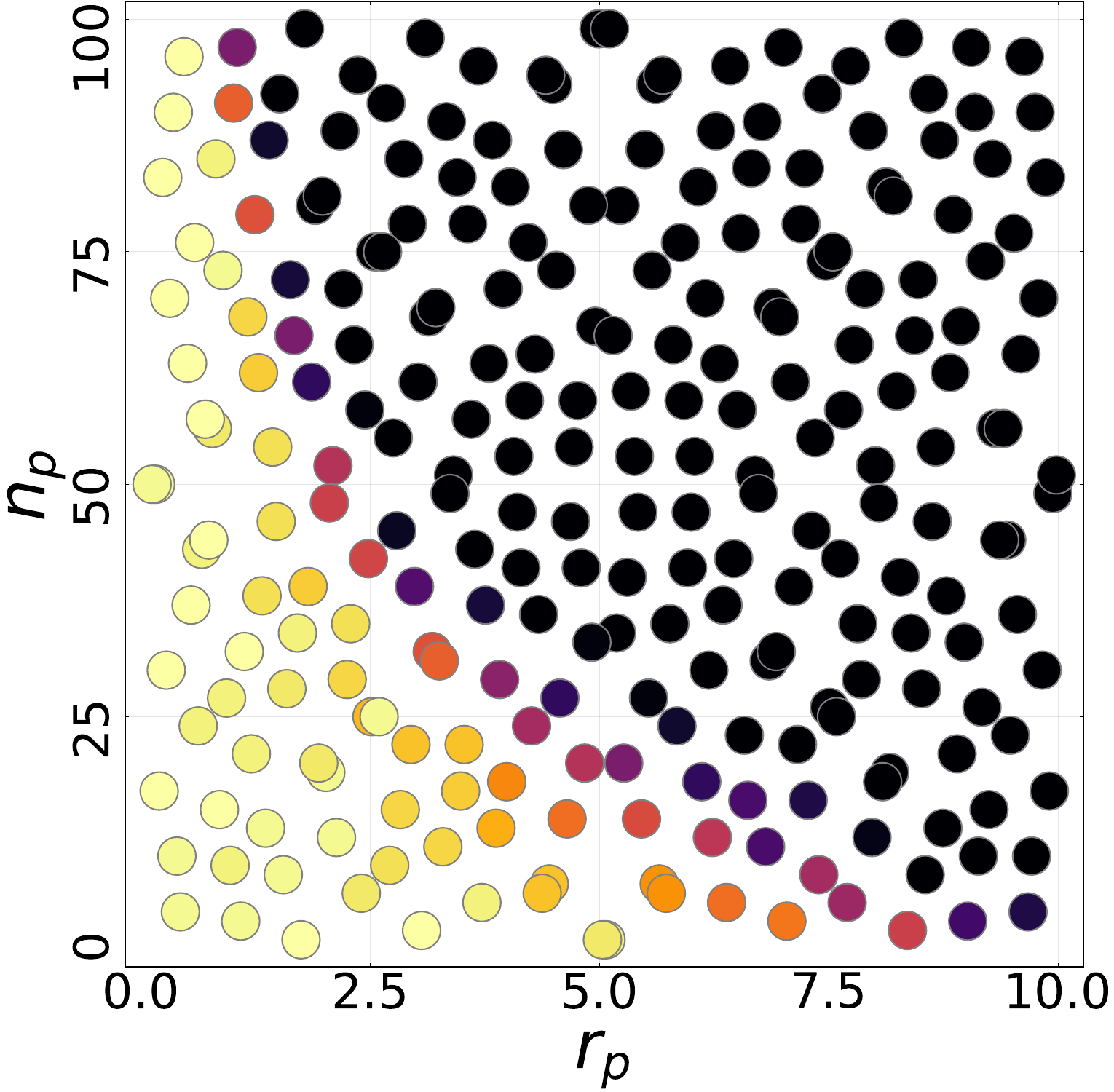}
         \caption{$\phi_B=30$}
     \end{subfigure}
     \hfill
     \begin{subfigure}[t]{0.188\linewidth}
         \centering
         \includegraphics[width=\textwidth]{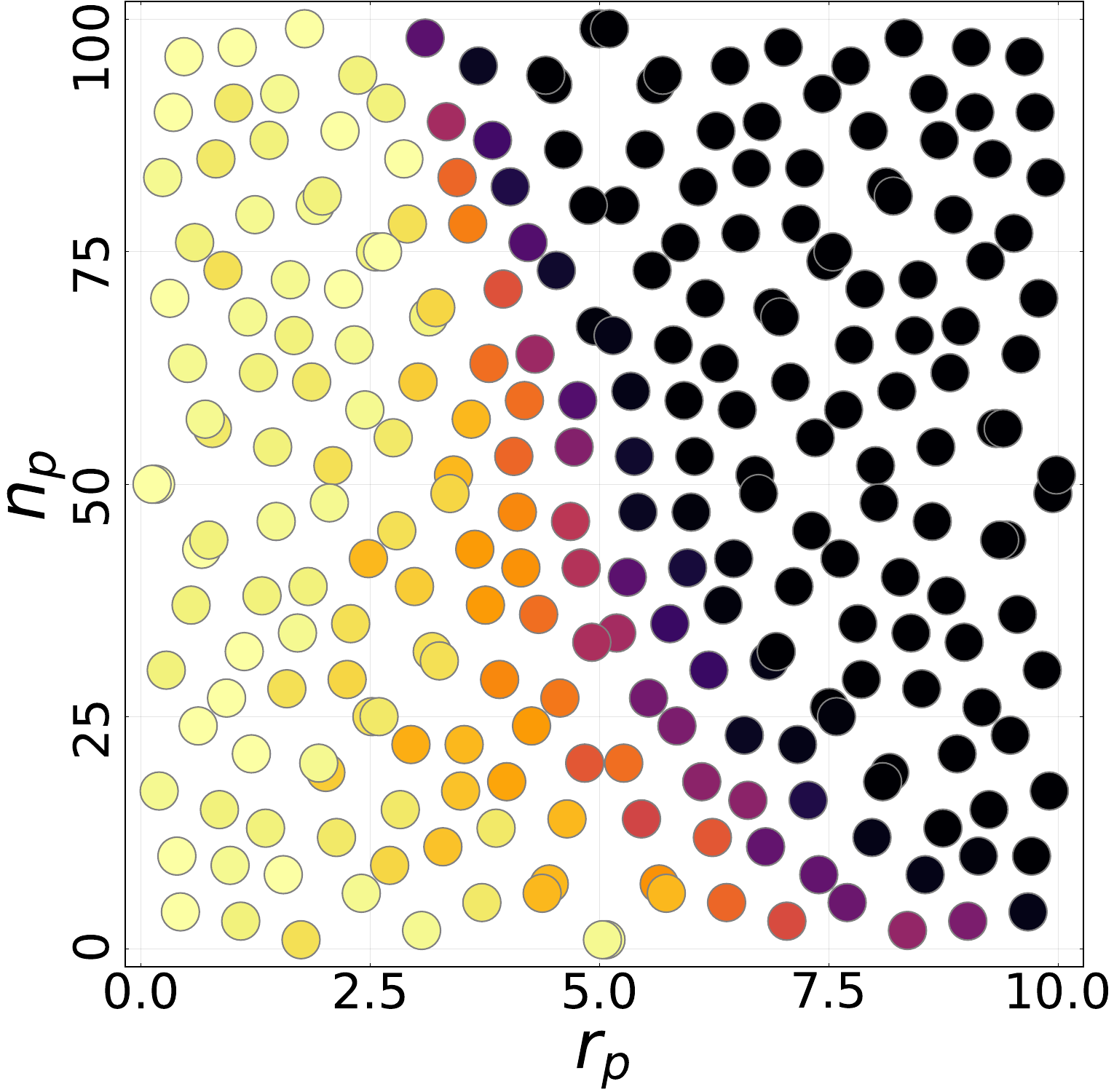}
         \caption{$\phi_B=100$}
     \end{subfigure}
     \hfill
     \begin{subfigure}[t]{0.222\linewidth}
         \centering
         \adjincludegraphics[width=\textwidth,trim={0 0 {.08\width} 0},clip]{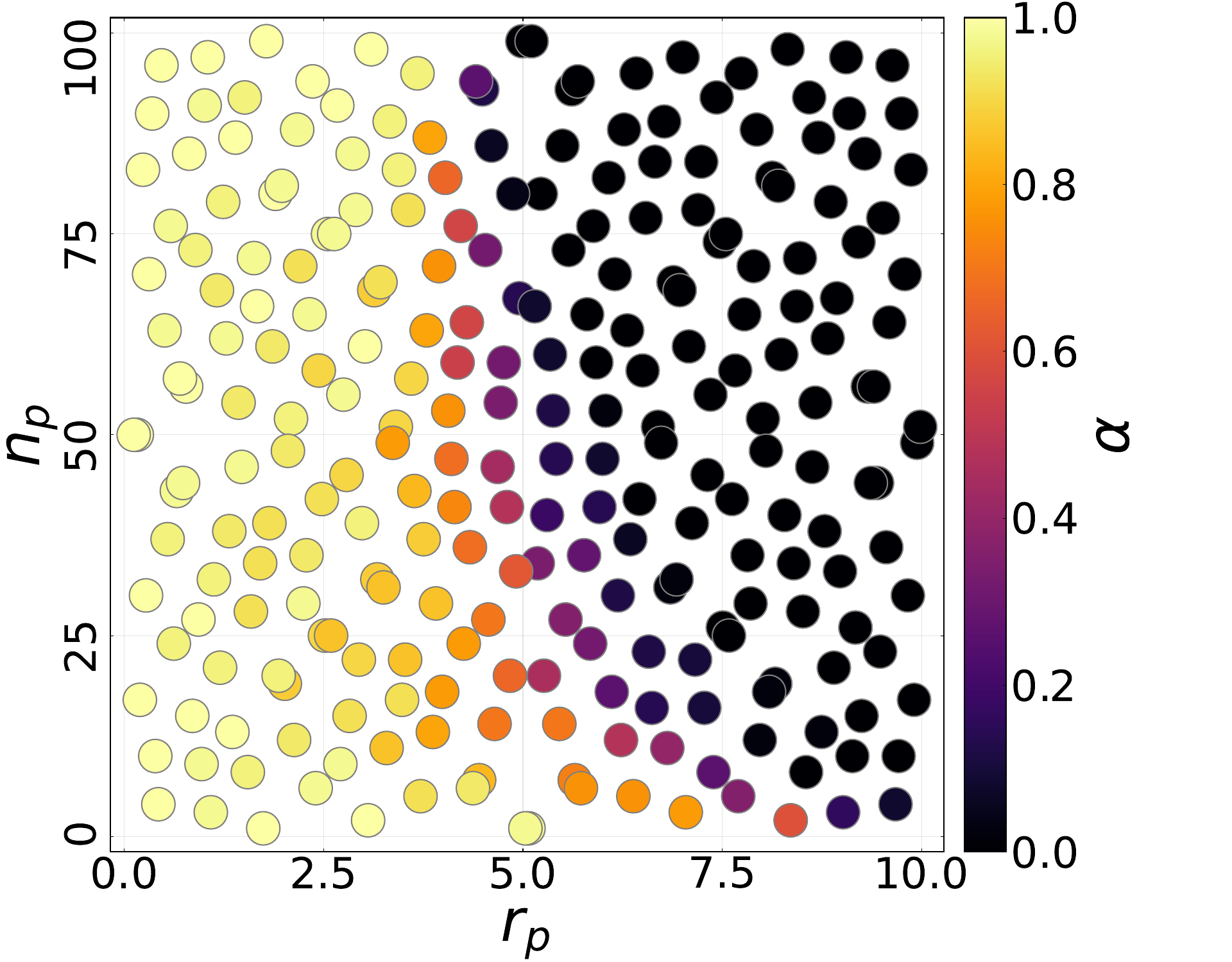}
         \caption{$\phi_B=300$}
         \label{max_battery_basin}
     \end{subfigure}
    
    \caption{Resilience basins for battery installation based on QMC sampling of the resilience assessment function $\alpha(r_p, n_p)$. The battery threshold parameter is fixed at $\lambda = 1$ and the battery budget $\phi_B$ is increased incrementally with each subfigure. The vertical axis indicates the number $n_p$ of appearing prosumers and the horizontal axis their ratio $r_p$ of power production to consumption.}
    \label{qmc_battery}
    
\end{sidewaysfigure}

\noindent This difference in maximal resilience is mirrored by the resilience basins in Figs. \ref{qmc_line} and \ref{qmc_battery}:
For both response options, the basins possess a distinct hyperbolic boundary which shifts away from the origin with increasing budget values ($\phi_L$ and $\phi_B$, respectively).
However, this shifting is where the differences between the response strategies lie:
First, the boundary seems to shift only sideways in Fig. \ref{qmc_battery} while it also shifts upwards in Fig. \ref{qmc_line}.
Second, the hyperbolic shape of the boundary line becomes steeper and almost vertical with increasing battery budget $\phi_B$ in Fig. \ref{qmc_battery}.
Together, this results in the fact that, for battery installation, the resilience basin is restricted to approximately the left half of the influence plane ($r_p \leq 5$), while line capacity upgrades achieve a complete covering of the plane.\\

\noindent The growth of the resilience basin with increasing budget is consistent with the budget dependency of the resilience measure itself: $R$ increases monotonically with increasing budget ($\phi_L$ and $\phi_B$, respectively). As can be seen in the log-log plots in Figs. \ref{line_budget_plot} and \ref{battery_budget_plot}, the curves have a distinct sigmoid shape in the majority of cases. The only exceptions are the curves for line upgrades with line allocation parameters of $\varepsilon \leq -4$.
The sigmoid shape means that the relative budget efficiency increases initially, but then decreases again as the curves plateau at their maximum value.

Two interesting characteristics of these sigmoid curves are the saturation point, which indicates the minimum budget necessary to achieve maximal resilience, and the turning point, indicating the point of maximal relative budget efficiency. The line budget necessary to achieve the maximum resilience value $R=1$ lies at $\phi_L \approx 100$ and the turning point of the sigmoid is located at $\phi_L \approx 1$. For batteries, the budget necessary for maximal resilience ($R\approx 0.7$) is $\phi_B \approx 100$, while the sigmoid has its turning point at $\phi_B \approx 10$.\\

\begin{figure*}
    \centering
    
    \begin{subfigure}[t]{0.488\textwidth}
        \vskip 0pt
        \centering
        \adjincludegraphics[width=\linewidth,trim={0 0 {.07\width} 0},clip]{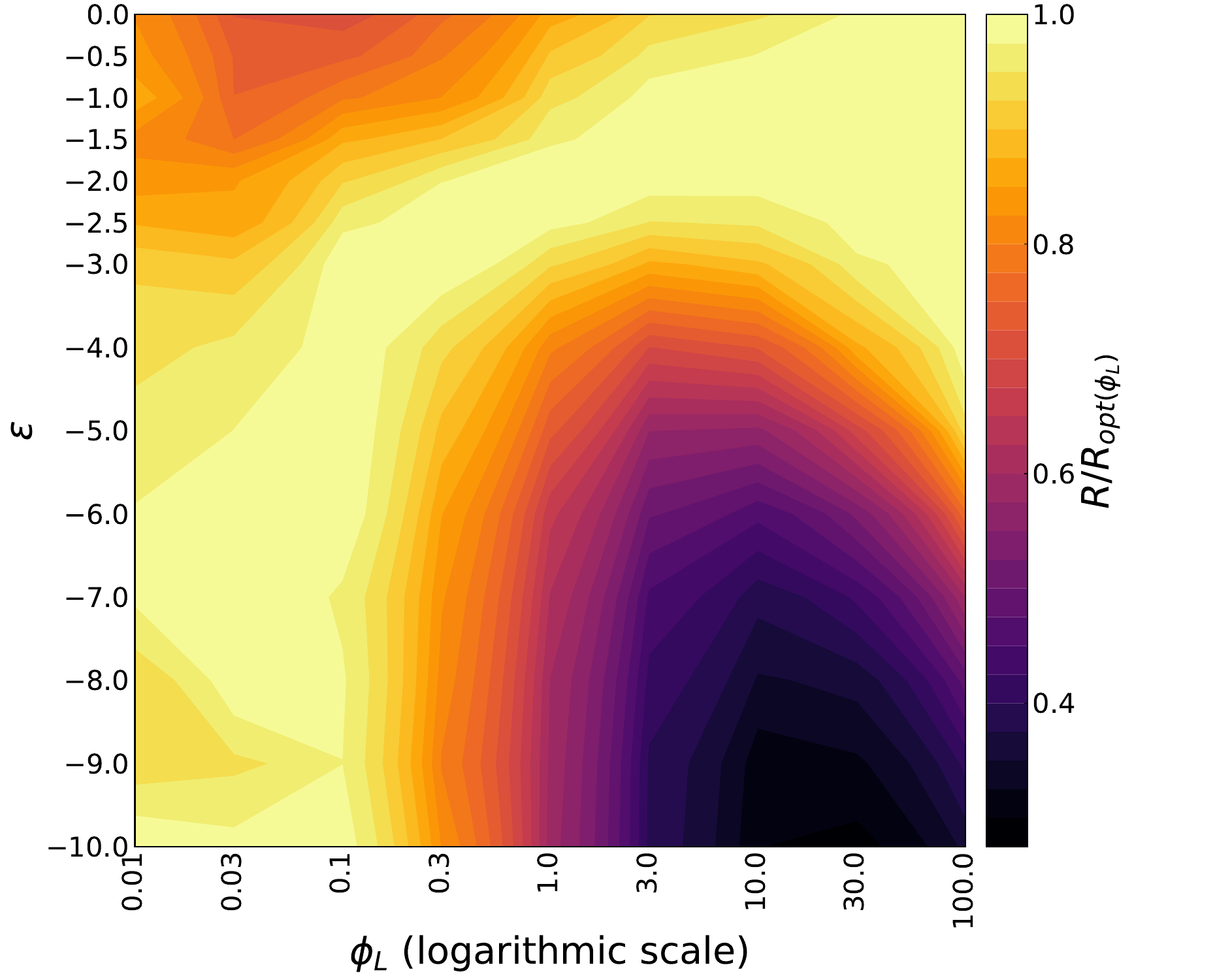}
        \caption{Optimal line allocation parameters $\varepsilon$ depending on the line budget $\phi_L$.}
        \label{line_optimal}
    \end{subfigure}
    \hfill
    \begin{subfigure}[t]{0.502\textwidth}
        \vskip 0pt
        \centering
        \adjincludegraphics[width=\linewidth,trim={0 0 {.07\width} 0},clip]{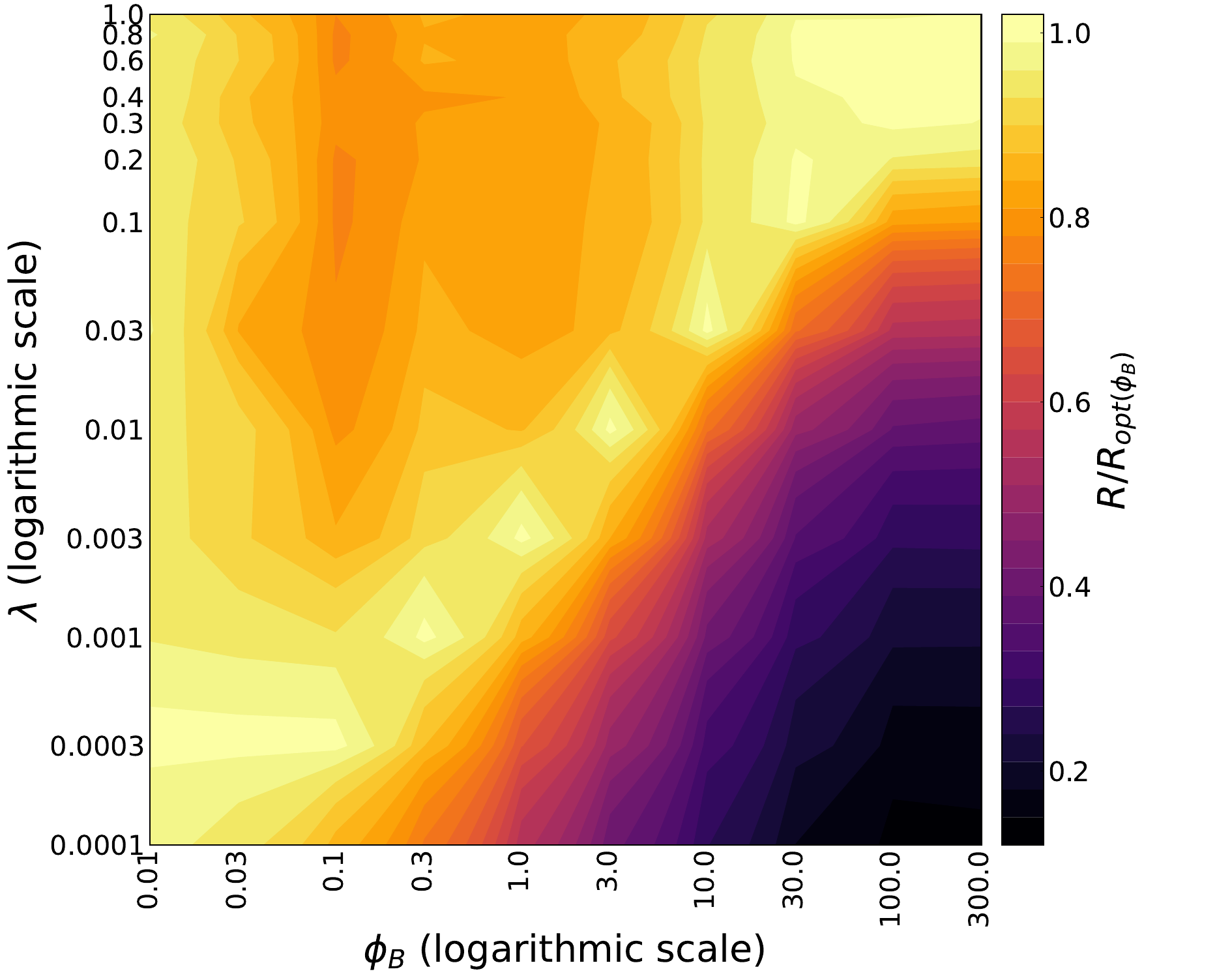}
        \caption{Optimal battery threshold parameters $\lambda$ depending on the battery budget $\phi_B$.}
        \label{battery_optimal}
    \end{subfigure}
    
    \caption{Illustration of the budget dependency of the optimal secondary response parameters for line upgrades and battery installation, respectively. The color indicates the resilience $R$ normalized to the optimal resilience $R_{\text{opt}}$ achieved at each budget value.}
\end{figure*}

\begin{figure*}
    \centering
    
    \begin{subfigure}[t]{0.466\textwidth}
        \centering
        \includegraphics[width=\linewidth]{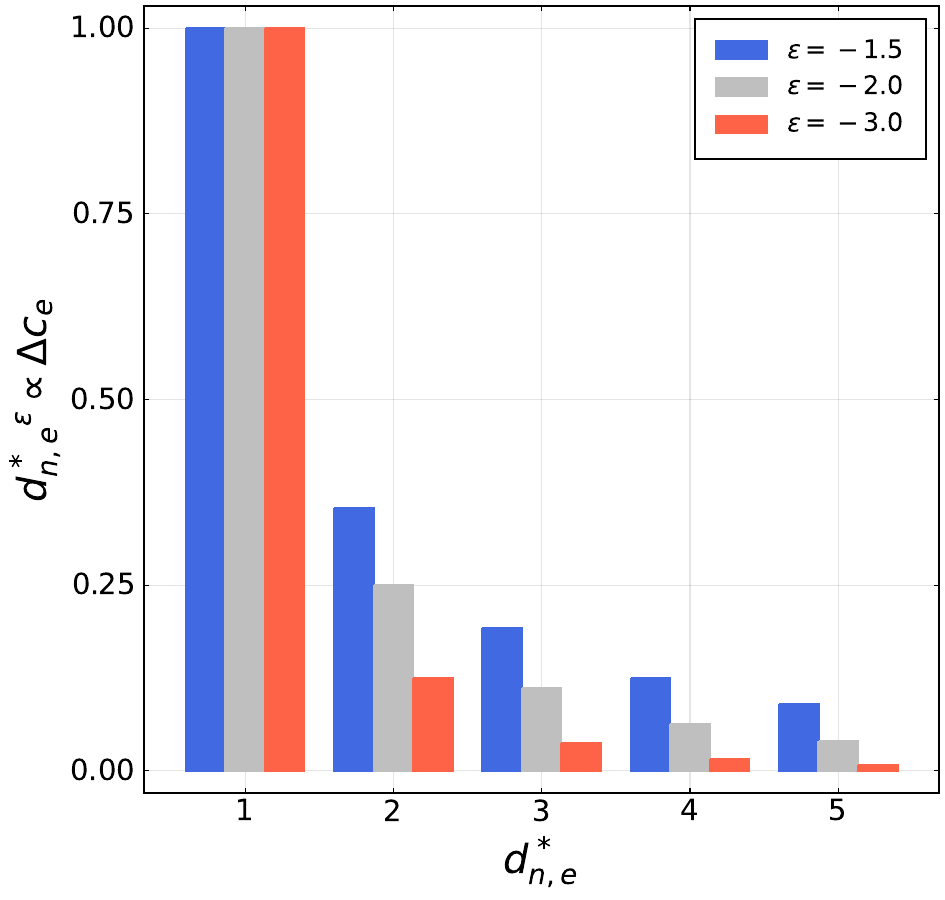}
        \caption{Line upgrade weight distributions up to the fifth neighborhood for different optimal values of the line allocation parameter $\varepsilon$.}
        \label{line_optimal_illustration}
    \end{subfigure}
    \hfill
    \begin{subfigure}[t]{0.524\textwidth}
        \centering
        \adjincludegraphics[width=\linewidth,trim={0 0 {.085\width} 0},clip]{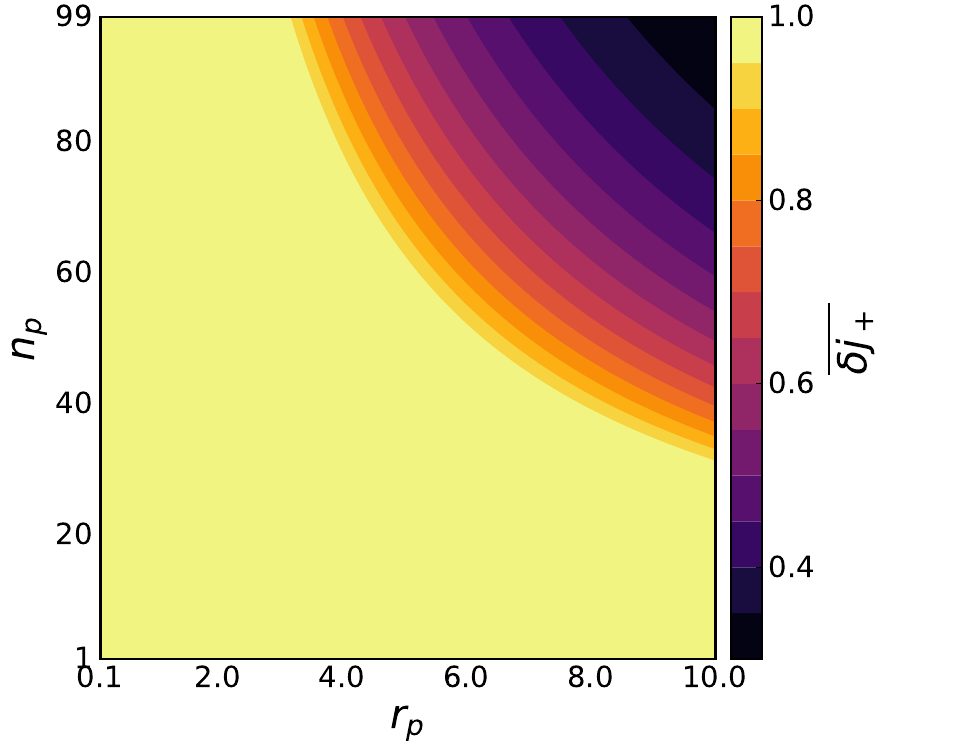}
        \caption{Relative excess injection $\overbar{\delta j_+}(r_p, n_p | \lambda=0.3)$ (see the Appendix for definition). The quantity in general reflects the fraction of power injection fluctuations that the batteries aim to absorb.
        This specific distribution at $\lambda=0.3$ represents the minimum fraction necessary to absorb in order to achieve optimal resilience using batteries.}
        \label{battery_threshold_illustration}
    \end{subfigure}
    
    \caption{Illustrations of selected values for the secondary response parameters $\varepsilon$ and $\lambda$, respectively. }
\end{figure*}

\noindent The secondary parameters ($\varepsilon$ and $\lambda$) both generally produce higher resilience at higher values, which can be seen in Figs. \ref{line_secondary_plot} and \ref{battery_secondary_plot}. In fact, their maximum values of $\varepsilon = 0$ and $\lambda = 1$, which correspond to the most simple response strategies (uniform line upgrades and complete elimination of fluctuations, respectively), are relatively safe choices, resulting in almost optimal resilience values across the range of budgets. However, for most budget values, optimal resilience $R$ is achieved with secondary parameters below their maximum:

With line upgrades, the optimum shifts from $\varepsilon \leq -3$ at low budget values to $\varepsilon = -2$ at $\phi_L = 1$ and then seems to settle at $\varepsilon = -1.5$ for $\phi_L \geq 3$. At very low budget values ($\phi_L \leq 0.03$), the height of the optimum vanishes compared to what seems to be random noise, and at very high budgets ($\phi_L \gtrsim 30$) the optimum becomes less sharp such that a broader range of allocation parameters can produce full resilience.
This budget dependency of the optimal allocation parameter is illustrated in Fig. \ref{line_optimal}.
Additionally, illustrations of the distance-dependent weights of the line upgrades for some values of $\varepsilon$ can be found in Fig. \ref{line_optimal_illustration}.

For the battery threshold parameter $\lambda$, the optimum is less pronounced and seems to steadily shift across orders of magnitude along with the battery budget $\phi_B$. As with line upgrades, the optimum diffuses at low budget values ($\phi_B \approx 0.01$) and becomes broader at high budget values ($\phi_B \geq 30$). However, the optimum does not become arbitrarily broad, but instead seems to be limited to the range of $0.3 \leq \lambda \leq 1$.
The influence dependency of the excess injection $\overbar{\delta j_+}(r_p, n_p)$ implied by this threshold value of $\lambda = 0.3$ is illustrated in Fig. \ref{battery_threshold_illustration}.
The budget dependency of the optimal value for $\lambda$ is illustrated in Fig. \ref{battery_optimal}.


\section{Discussion}
\label{sec:discussion}

The first application of our framework to the single-node system is not meant to derive any new insights. The system has been studied extensively, and our chosen formulation of the resilience measure in equation \eqref{eq:single_node_res} is essentially equivalent to finite-time basin stability \citep{schultz2018bounding}, with the finite recovery time translating to a finite cost.
Yet, we believe that this example proves the generality of our framework.\\

\noindent Regarding the second application, the framework was able to confirm most of our intuitive expectations about the response strategies of the power grid.

The fact that battery installation has an inherent limitation in regard to the maximum achievable resilience $R$ was not foreseen, but actually makes intuitive sense since, after all, the batteries can only eliminate fluctuations in the power injections.\\

\noindent Going into the details, the hyperbolic shape of the resilience basins was not expected, but can be understood with the following consideration:

The overall average power supply $\Sigma\, \overbar{s}$ provided by prosumers is linearly related to both of the adverse influence parameters: $\Sigma\, \overbar{s} \propto n_p \times r_p$, which can be rearranged as \mbox{$n_p(r_p) \propto \Sigma\, \overbar{s} \times (r_p)^{-1}$}. Each hyperbola, therefore, represents a line of constant average overall power supply $\Sigma\, \overbar{s} = \text{const}$, and one of those hyperbolae indicates the maximum supply that the power grid can sustain without implementing response options, solely by means of the initial power line capacities.

Power line upgrades then seem to be able to indefinitely and linearly increase the maximal supply which the power grid can cope with. For example, Fig. \ref{qmc_line_net_neutral}, where $\phi_L = 3$, shows a boundary around $n_p \times r_p = 99$ which corresponds to the power grid being net-neutral, meaning the average power $\sum \overbar{s}$ supplied by the prosumers equals the average demand $\sum \overbar{d}$ by all nodes.

Battery installation, on the other hand, cannot increase the average supply tolerance, only the fluctuation tolerance. The average magnitude of any local fluctuations is only proportional to $r_p$, not to $n_p$, which can explain why the basin's hyperbolic boundary only shifts to the right in Fig. \ref{qmc_battery}. This boundary shift can, however, only be sustained to the point where the average power injections exceed the tolerance of the initial line capacities, which seems to happen at $r_p \approx 4$.\\

\noindent The monotonous budget dependency of $R$ seen in Figs. \ref{line_budget_plot} and \ref{battery_budget_plot} is in line with intuitive expectations.
The exact sigmoid behavior was not predicted, but can be partially understood by considering the shape of the resilience basins in combination with the definition of the influence density function $\varrho(\vec{i})$:

The first, upward-curving half of the sigmoid suggests that each budget parameter ($\phi_L$ and $\phi_B$, respectively) increases the area of the resilience basin exponentially or as a power function rather than linearly. The turning point and subsequent plateauing of the sigmoid can be explained the following way: First, the area of the resilience basins is inherently limited to the size of the chosen adverse influence plane. Second, at low budgets, the resilience basins are located close to the origin of the influence plane, where the probability density $\varrho(\vec{i})$ was defined to be the greatest. With increasing budgets, the basins grow into regions of lower and lower probability density, decreasing the gain in probability with each budget increment.\\

\noindent The general dependency of $R$ on the line allocation parameter $\varepsilon$ matches the expectations: The optimal strategy is a value of $\varepsilon<1$ for all budgets. However, the optimum is not independent of the budget, as was hypothesized. Instead, it is monotonically and non-linearly increasing with the budget $\phi_L$. This means that, at lower budgets, it is best to prioritize lines closer to the prosumers. The exact optimal values of $\varepsilon$ and their budget dependency are not obvious to justify, but it might be possible to relate them to the average node degree of the network or its small-worldness.\\

\noindent The battery threshold parameter dependency $R(\lambda)$ also matched our expectations. The budget-dependent shift of the optimal value of $\lambda$ is mostly linear, as predicted. This confirms that the realizable amount of excess injection (which is proportional to $\lambda$) can be scaled up linearly by increasing the battery sizes (which are proportional $\phi_B$).

However, the optima of $\lambda$ in Fig. \ref{battery_secondary_plot} are more shallow than expected. The existence of the optima was hypothesized based on a sweet spot between absorbing as many fluctuations as possible and preventing battery clogging or emptying. The fact that that the optimal resilience measure $R(\lambda_\text{opt})$ is only slightly higher than $R(\lambda=1)$ suggests that the risk of battery clogging is significantly lower than the risks posed by allowing greater injections, even at small battery sizes.

The threshold value for the battery threshold parameter found as $\lambda \approx 0.3$ seems intriguing at first, but may have a less exciting explanation: Fig. \ref{battery_threshold_illustration} shows the so-called relative excess injection $\overbar{\delta j_+}(r_p, n_p)$ (see the Appendix for details) corresponding to the case of $\lambda=0.3$\,.
This quantity reflects the fraction of power injection deviations that the batteries aim to absorb.
One can see that the only adverse influences $\vec{i}$ for which $\overbar{\delta j_+}(\vec{i})<1$ are located in the top right corner of the influence plane, at $r_p \gtrsim 3.5$ and $n_p \gtrsim 35$.
Coincidentally, this region almost perfectly touches the boundary of the largest possible resilience basin for batteries seen in Fig. \ref{max_battery_basin}. This means that, for adverse influences within that maximal basin, which are the hard limit for what batteries can provide resilience against, $\lambda \gtrsim 0.3$ has the same effect as $\lambda = 1$. $\lambda = 0.3$ just happens to be the smallest value for which the detrimental zone of $\overbar{\delta j_+} < 1$ does not chip away at the basin. At lower values of $\lambda$, however, the hyperbolic zone of $\overbar{\delta j_+} < 1$ overlaps with the maximum resilience basin, reducing the effective basin size.

\section{Conclusion}
\label{sec:conclusion}

Assessing both the well-studied single-node dynamical system and a unique probabilistic power grid
serves as a thorough proof of concept for our resilience measurement approach. We believe that our framework will be able to provide useful insights throughout a wide range of applications.
Further, the scenario of prosumers destabilizing the power grid does not only represent a highly relevant scenario in the context of climate change mitigation, but also a great example of how resilience can be studied in the context of adverse side effects caused by otherwise desirable processes.

The key feature of our framework is that the resilience basins do not necessarily exist within the phase space of the system (as in basin stability), but in a separately defined space of adverse influences. Likewise, the space of response options is separate and of arbitrary dimension.
We believe that this generally requires more conscious effort and subjective decisions by the modeler, but has the potential to produce more concrete and meaningful results, as well as allowing application to non-dynamical systems that are not described by differential equations.

Since the framework does not require exact knowledge of the system's state, only of its sustainant, one application scenario might be neural networks which are trained to mimic a highly complex system's sustainant behavior depending on different adverse influences. Further, since we interpret control theory as the study of adaptive resilience, resilience basins may be a useful tool in quantifying the success of control-theoretical mechanisms.\\

\noindent Regarding the application to the probabilistic power grid, it has to be clearly stated that the simplicity of our model does not allow for direct quantitative inferences about real-world power grids.
The model makes many specific assumptions about the actual grid dynamics. It assumes direct current, lossless linear flows without cycles, stochastic injections which are not seasonally coherent, static injections over a duration of 1 minute, equal PV sizes for all prosumers, immediate blackouts in case of power deficiency, automatic power line reboots, unregulated PV installation, and vastly simplified battery functionality. It further excludes the conversion of material costs to financial costs while also ignoring construction costs of line upgrades, installation costs of batteries as well as the time scales on which both response options are performed.

Keeping this in mind, for our model it was found that the battery installation response is inherently limited in terms of providing resilience, while power line capacity upgrades displayed unlimited resilience potential. Metaphorically speaking, batteries act like a wave breaker, which can dissipate high waves in stormy weather, but cannot prevent rising sea levels from flooding the shores. Power line upgrades on the other hand act like digging out the sea floor, allowing the ocean volume to grow without raising the sea level.
These qualitative differences are nothing revelatory, but our framework is uniquely able to put a number to them by quantifying the maximal resilience effect of both response strategies (0.7 vs. 1), at least in the specific prosumer scenario that was chosen.

Further, the framework was able to provide insight into budget efficiency in several ways: First, the general budget dependency was found to be sigmoidal and therefore to have both a turning point and a saturation point. Knowledge of the saturation point prevents wasting budget by overly adapting or transforming the power grid when aiming for complete resilience. On the other hand, if only partial resilience $R<1$ is deemed sufficient, the turning points present themselves as optimal budget choices.

Second, the framework identified optimal secondary parameters for both response strategies and how they depend on the available budget. However, the relative advantage of these optimal strategies was relatively small. The most simple strategies (uniformly allocated line upgrades and full absorption of injection fluctuations) proved to be consistently effective alternatives, coming close to maximal resilience at all budgets.

This is relevant because those simple strategies are independent of the number of prosumers and their PV size, making them potentially feasible in a completely unregulated scenario.
Particularly, uniform line upgrades are even independent of the prosumer locations, which entirely eliminates the need to \emph{react} to their emergence. Therefore, the power lines could instead be upgraded prophylactically and thereby the grid would be made \emph{robust} instead of resilient.
Whether the trade-off between this independence and the suboptimal budget efficiency is desirable, however, will depend on additional considerations.\\

\noindent Considering real-world conditions, the installation of batteries in prosumer households is arguably simpler to realize than upgrading all power lines, especially if the latter are built underground. Therefore, a combination of both response strategies is probably most feasible: Installing batteries that are large enough to absorb all fluctuations, but also upgrading selected power lines to be able to cope with the increased average power injections by prosumers. Optimizing this combined strategy for all instances of adverse influences will require a higher-dimensional analysis considering the parameters of both response options simultaneously.

It can further be noted that both of the examined response strategies (in general, not only the simple cases mentioned above) actually have the potential to provide robustness instead of resilience: If the installation of PV by prosumers simply were to be regulated, the responses could be implemented before the PV modules are actually connected to the power grid. This principle may actually apply to social-technological systems in general due to the innate ability of its agents to communicate.\\

\noindent Drawing an analogy to the classification of resilience based on a system's response options by \citet{barfuss}, one may be able to classify adverse influences based on their temporal shape. Staying in the context of power grids, one could differentiate between \emph{temporary changes} in energy production vs. \emph{permanent changes} in energy production vs. \emph{permanent risk of permanent changes} in energy production. Translating these categories of adverse influences
to dynamical systems in general, one might classify state perturbations as\emph{momentary} external force pulses vs. \emph{constant} external forces vs. \emph{permanently variable} external forces.
These three categories may even correspond somewhat to persistence, adaptation and transformation as the responses necessary to cope with them.

Further, \citet{barfuss} analyze resilience from a multi-agent-environment perspective. They introduce the concepts of general vs. specific resilience (resilience of the whole social-technological system vs. of a sub-system, for example the technological sub-system), as well as first-order vs. second-order resilience (resilience for single agents vs. resilience that mutually benefits other agents).
In this regard, one could say that the power grid model examined in this paper had multiple independent agents driving the adverse influence (consumers turning into prosumers) and a single agent managing the response options (the grid operator upgrading lines or installing batteries).

This aspect could be potentially expanded upon by considering multi-agent adverse influences that are \emph{dependent} on each other (e.g. consumers copying their neighbor's decision to install PV if they perceive it to be advantageous) as well as multi-agent responses (e.g. prosumers copying their neighbor's battery installation decision or even sharing their battery capacities with one another).

\newpage
\begin{acknowledgments}
We all want to thank Frank Hellmann, leader of the ICoNe workgroup at PIK, for the fruitful discussions that eventually lead to this paper.
S.B. further thanks ICoNe members Mehrnaz Anvari and Anton Plietzsch for their support during his Bachelor's thesis preceding this paper.

We are immensely grateful for the power data resources provided by Elisavet Proedrou of the NOVAREF project at the German Aerospace Center (DLR) as well as those made available via the Open Power Systems Data (OPSD) project.

This work was funded by the Deutsche Forschungsgemeinschaft (DFG, German Research Foundation) – KU 837/39-1 / RA 516/13-1 and by the Leibniz Association (project DOMINOES). 

J.F.D. is grateful for financial support by the European Research Council advanced grant project ERA (Earth Resilience in the Anthropocene; grant ERC-2016-ADG-743080) and the German Federal Ministry for Education and Research (BMBF) through the 'PIK Change' framework (Grant No. 01LS2001A).

All authors gratefully acknowledge the European Regional Development Fund (ERDF), the German Federal Ministry of Education and Research and the Land Brandenburg for supporting this project by providing resources on the high performance computer system at the Potsdam Institute for Climate Impact Research (PIK).

\end{acknowledgments}

\section*{Data and Code Availability Statement}

The single-node model is implemented using the programming language Python in version 3.9.16\,.
The power grid model is implemented using the programming language Julia \citep{julia} in version 1.5.3\,.
The manually filtered datasets as well as the model code are available from the corresponding author upon request.

\clearpage
\appendix

\section{Details: Probabilistic Power Grid Model}

\subsection{Network Structure}

The nodes are embedded in the two-dimensional Euclidean plane and their coordinates are drawn randomly from a uniform distribution. Connections are made according to attachment rules controlled by a small set of parameters. 
The parameter values used in the growth algorithm were chosen empirically to make the graphs resemble real-world low-voltage power grids, matching the scenario of small-scale decentralized PV power generation by prosumers. The chosen parameter values and their roles are summarized in Table \ref{tab:synth_net}.

\begin{table}[h]
    \centering
    \begin{tabular}{llp{5cm}}
         Parameter & Value & Role \\
         \hline
         $N$ & 100 & Total number of nodes \\
         $N_0$ & 15 & Initial number of nodes, connected by minimum spanning tree \\
         $s$ & 0 & Probability of creating a new node by splitting a random existing line in half (vs. creating one in a random location and linking it to its closest neighbor) \\
         $p$ & 0.15 & Probability of linking a newly created node to the network via a second, redundant line \\
         $q$ & 0.15 & Probability of linking an existing node via a second, redundant line \\
         $r$ & 0.2 & An exponent that determines which lines are preferred when creating redundant lines\\
         \hline
    \end{tabular}
    \caption{Parameter roles in the network growth algorithm by \citet{synth_net} and the values used in our power grid model.}
    \label{tab:synth_net}
\end{table}





\subsection{Power data sources}

The power consumption data stem from the NOVAREF project \citep{novaref} of the German Aerospace Center (DLR). It was kindly provided by Elisavet Proedrou at the DLR Institute of Networked Energy Systems in Oldenburg, Germany. It represents the averaged load profile of 12 residential houses in Oldenburg, Lower Saxony, measured throughout the year 2013.

The PV power generation data set is part of a larger publicly available data collection by Open Power System Data \citep{opsd}. It consists of the solar irradiance measured by a suburban residential house's rooftop PV system in the south of Germany, covering a time span from 2015 to 2017. The irradiance measurements are treated as power measurements.

\subsection{Initial Power Line Capacities}

The optimization of the initial power line capacities for centralized supply is achieved by generating random power demand time series where all nodes are treated as regular consumers.

Here, however, the time series are not coherently generated from daily chunks, but instead only from a random subset of their time steps (which are still aligned across the nodes regarding time of day). This can be thought of as a collection of snapshots from a longer, coherent time series. This way, it takes less time steps to capture the variability of the power consumption, reducing the computing time.

From this snapshot time series, the power flows are calculated and, for each edge, the maximum occurring flow is extracted. The absolute values of these maximum flows are then multiplied with a safety factor to produce the line capacities. The length of the snapshot time series was chosen as 1000 time steps and the safety factor as 1.75\,.

\subsection{Response options}

\subsubsection{Power Line Upgrade Allocation}

For the purpose of line upgrade allocation, the shortest path distance $d^*$ of an edge to a node is defined as follows:
\begin{equation}
    d^*_{e,n} = \min\{ d_{\text{src}(e),n}, d_{\text{dst}(e),n} \} +1,
\end{equation}
where $n$ is a node index, $e$ is an edge index, $d_{i,j}$ is the shortest path distance between two nodes $i$ and $j$, and $\text{src}(e)$ and $\text{dst}(e)$ are the incident nodes of edge $e$.

To make the capacity upgrades proportional to the weighted distances to the prosumers, the adaptation budget portion allocated to each line also has to be proportional to the length $l$ of the line itself. Therefore, the capacity upgrades $\Delta c$ are determined by the following formula:
\begin{equation}
    \Delta c_e = \frac{1}{l_e} \times \beta_L \times \frac{
    \sum_{n \in U_p} l_e \times {d^*_{e,n}}^\varepsilon}{\sum_f \sum_{n \in U_p} l_f \times {d^*_{f, n}}^\varepsilon},
\end{equation}
where $e$ and $f$ are edge indices, $n$ is a node index, $U_p$ is the set of all prosumer node indices, and $\varepsilon$ is the line allocation parameter.

\subsubsection{Battery Mechanism}

Since all consumer and prosumer nodes have an identical average demand of  $\overbar{d} = 1\,$p.u., the reference value $E_0$ for the battery budget is calculated as follows:
\begin{equation}
    E_0 = 99 \times 1\,\text{p.u.} \times 1\,\text{d} = 142,560\,\text{p.u.}\times\text{s},
\end{equation}

Without batteries, the power injection $j$ at each node simply equals the difference between the node's power supply $s$ and its demand $d$. The batteries convert $j$ into a modified injection $j^*$ in a very rudimentary way by limiting how far it can deviate from the mean injection $\overbar{j}$. Any deviation exceeding the limits will be absorbed by the battery such that the remaining injection equals the limit value, if possible. This, of course, depends on the battery size $b_{\text{max}}$ and the battery's energy content $b$ from the previous time step:
\begin{equation}
\begin{gathered}
    \Delta b_n(t) = \left\{ \begin{array}{l}
        \min \big\{ j_n(t)-j_{\text{max}}, b_{n, \text{max}}-b_n(t-1) \big\}\\[2pt]
        \hspace{100pt} \text{if $j_n(t)>j_{\text{max}}$} \\[5pt]
        \min \big\{ j_n(t)-j_{\text{min}}, b_n(t-1) \big\}\\[2pt]
        \hspace{100pt} \text{if $j_n(t)<j_{\text{min}}$} \\[5pt]
        0 \hspace{95pt} \text{else,}
        \end{array} \right.
    \\[5pt]
    j_n^*(t) = j_n(t) - \Delta b_n(t),
\end{gathered}
\end{equation}
where $\Delta b$ is the battery charge/discharge, $n$ is a node index, $t$ is a time step index, and $j_{\text{min}}$ and $j_{\text{max}}$ are the limits of allowed injection values.

After having installed the batteries at $t=t_4$, they are initiated at 50\,\% charge so that they can absorb injection deviations in both directions equally well:
\begin{equation}
    b_n(t_4) = 0.5 \times b_{n, \text{max}}.
\end{equation}

\begin{figure}
    \centering
    \includegraphics[width=\linewidth]{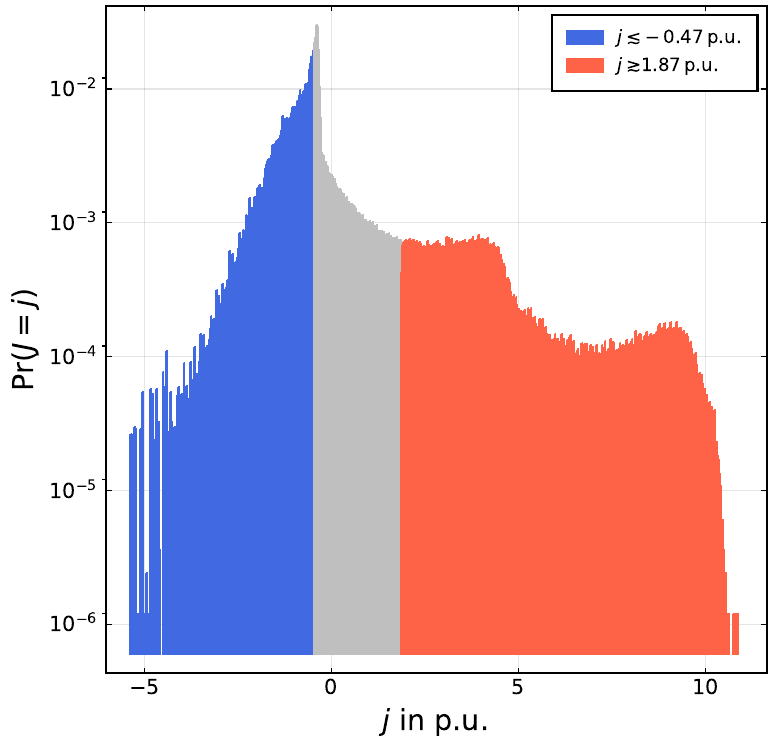}
    \caption{Logarithmic histogram of the distribution of possible injection values $j$ for a prosumer ratio of $r_p=1$. The mean value is $\overbar{j}=(r_p-1)$\,p.u. $=0$. The red columns indicate the excess above the limit $j_{\text{max}} \approx 1.87$\,p.u. which is chosen such that $\overbar{\delta j_+}(j_{\text{max}}) = 0.5$\,. The blue columns are the opposite excess below $j_{\text{min}} \approx -0.47$\,p.u. which has to be absorbed to preserve the mean injection value.}
    \label{histogram}
\end{figure}

The distribution of all possible power injections is generated from the fixed distribution of power demand ($\overbar{d} = 1$\,p.u.) and the variable distribution of power supply ($\overbar{s} = r_p$\,p.u.). An exemplary combined distribution for the case $r_p=1$ is illustrated as a logarithmic histogram in Fig. \ref{histogram}. As evident, it is not symmetrical around its mean and, therefore, the upper and lower deviation limits have to be different to preserve the mean $\overbar{j^*} = \overbar{j} =(r_p-1)$\,p.u..

It is decided to control the maximum injection $j_{\text{max}}$ using the secondary battery response parameter, and determine the minimal injection $j_{\text{min}}$ such that the mean injection $\overbar{j^*} = \overbar{j}$ is preserved.
To determine a sensible value for $j_{\text{max}}$, we introduce a quantity which we call the \emph{excess injection} $\overbar{\Delta j_+}$ and which is dependent on $j_{\text{max}}$. $\overbar{\Delta j_+}$ is defined as the sum of all additional injections above $j_{\text{max}}$, weighted by their probability of occurrence:
\begin{equation}
    \overbar{\Delta j_+}(j_{\text{max}}) = \sum_{j>j_{\text{max}}} (j-j_{\text{max}}) \times \text{Pr}(J=j),
\end{equation}
where $J$ is the random variable from which the prosumers' power injections are drawn. $\overbar{\Delta j_+}$ is deliberately not normalized to reflect the difference in probability of excess injection depending on the limit value $j_{\text{max}}$. That way, $\overbar{\Delta j_+}$ reflects the amount of energy that is redirected from the PV supply to the battery periodically. 
The maximum value for this excess quantity is achieved when $j_{\text{max}} = \overbar{j}$, meaning when all deviations above the mean are eliminated. For easier notation, we define the \emph{relative excess injection} $\overbar{\delta j_+}$:
\begin{equation}
    \overbar{\delta j_+}(j_{\text{max}}) = \frac{\overbar{\Delta j_+}(j_{\text{max}})}{\overbar{\Delta j_+}(\overbar{j})}.
\end{equation}

The secondary battery response parameter, called the \emph{battery threshold parameter} $\lambda$, now proportionally controls the value of this relative excess which the batteries absorb. However, besides $\lambda$, both influence parameters $r_p$ and $n_p$ are considered as well:
Increasing $r_p$ means increasing the variance of the distribution of possible values for $j$ and thereby the maximum excess injection. This dependency is empirically found to be approximately $\overbar{\Delta j_+}(\overbar{j}) \propto r_p$. It, therefore, makes sense to choose $\overbar{\delta j_+}(j_{\text{max}}) \propto \frac{1}{r_p}$.

Further, increasing $n_p$ reduces the size of each individual battery (given a fixed budget) and, therefore, lessens the amount of fluctuations it can absorb. For this reason, the relative excess is chosen to have a dependency $\overbar{\delta j_+}(j_{\text{max}}) \propto \frac{1}{n_p}$.

Combining these dependencies, we get an implicit formula that defines the injection limit $j_{\text{max}}$ via
\begin{equation}
    \overbar{\delta j_+}(j_{\text{max}}) = \min\left\{1, \frac{\lambda}{\frac{r_p}{10} \times \frac{n_p}{99}}\right\}
    \label{j_max},
\end{equation}
where the minimum function ensures that $\overbar{\delta j_+} \leq 1$. The additional factors in the denominator are chosen such that $\lambda$ equals the minimum relative excess which is absorbed in the case of the most extreme influence ($r_p=10, n_p=99$).

As mentioned earlier, the lower injection limit $j_{\text{min}}$ is now defined by the implicit condition that the altered injection $j^*$ maintains the same mean value as the unaltered injection $j$:
\begin{equation}
\overbar{j^*}(j_{\text{min}}, j_{\text{max}}) = \overbar{j} = (r_p-1)\,\text{p.u.}.
\end{equation}

\clearpage
\bibliography{sources}

\end{document}